\documentclass[final,11pt]{article}

\usepackage{multirow}
\usepackage[nottoc]{tocbibind}
\usepackage{geometry}
\usepackage{subcaption}
\usepackage{placeins}
\usepackage[affil-it, auth-sc]{authblk}
\usepackage[english]{babel}
\usepackage[utf8]{inputenc}
\usepackage{graphicx}
\usepackage{amssymb}
\usepackage{amsthm}
\usepackage{amsmath}
\usepackage{amsfonts}
\usepackage{verbatim}
\usepackage{lineno}
\usepackage{todonotes}
\presetkeys{todonotes}{inline, color=blue!30, size=\small}{}

\usepackage{color, soul}
\definecolor{darkred}{rgb}{0.9, 0.0, 0.0}
\definecolor{darkgreen}{rgb}{0.0, 0.5, 0.0}

\usepackage[colorlinks,bookmarks,linkcolor=darkred, citecolor=darkgreen]{hyperref}
\usepackage{eso-pic}
\usepackage[sort&compress,numbers]{natbib}
\usepackage{doi}
\usepackage{paralist,epsfig}
\usepackage{relsize}
\usepackage{nicefrac,esint}
\usepackage{float}
\usepackage{cleveref}

\def\slash#1{#1\!\!\!/\!\,\,}

\begin{document}

\AddToShipoutPictureFG*{
    \AtPageUpperLeft{\put(-60,-60){\makebox[\paperwidth][r]{LA-UR-22-29964}}}  
    \AtPageUpperLeft{\put(-60,-74){\makebox[\paperwidth][r]{FERMILAB-PUB-22-900-T}}}  
    }
    
\title{Radiative corrections to inverse muon decay for accelerator neutrinos}
\author[1]{Oleksandr~Tomalak \thanks{tomalak@lanl.gov}}
\affil[1]{Theoretical Division, Los Alamos National Laboratory, Los Alamos, NM 87545, USA \vspace{1.2mm}}
\author[2,3]{Kaushik Borah}
\affil[2]{University of Kentucky, Department of Physics and Astronomy, Lexington, KY 40506, USA \vspace{1.2mm}}
\affil[3]{Fermilab, Theoretical Physics Department, Batavia, IL 60510, USA \vspace{1.2mm}}
\author[2,3]{Richard~J.~Hill}
\author[4]{Kevin~S.~McFarland}
\affil[4]{University of Rochester, Department of Physics and Astronomy, Rochester, NY 14627, USA \vspace{1.2mm}}
\author[4]{Daniel~Ruterbories}

\date{\today}

\maketitle

Inverse muon decay ($\nu_\mu e^- \to  \nu_e \mu^-$) is a promising tool to constrain neutrino fluxes with energies $E_{\nu} \ge 10.9~\mathrm{GeV}$. Radiative corrections introduce percent-level distortions to energy spectra of outgoing muons and depend on experimental details. In this paper, we calculate radiative corrections to the scattering processes $\nu_\mu e^- \to  \nu_e \mu^-$ and $\bar{\nu}_e e^- \to  \bar{\nu}_\mu \mu^-$. We present the muon energy spectrum for both channels, double-differential distributions in muon energy and muon scattering angle and in photon energy and photon scattering angle, and the photon energy spectrum for the dominant $\nu_\mu e^- \to  \nu_e \mu^-$ process. Our results clarify and extend the region of applicability of previous results in the literature for the double differential distribution in muon energy and photon energy, and in the muon energy spectrum with a radiated photon above a threshold energy. We provide analytic expressions for single, double and triple differential cross sections, and discuss how radiative corrections modify experimentally interesting observable distributions.
\tableofcontents

\section{Introduction}
\label{sec1}

Scattering of neutrino beams from atomic electrons provides us a ``standard candle" for constraints on the neutrino fluxes at accelerator-based experiments. For example, the MINERvA experiment exploits the elastic scattering channel $\nu_\ell e^- \to \nu_\ell e^-$~\cite{MINERvA:2015nqi,MINERvA:2016iqn,MINERvA:2019hhc,MINERvA:2022vmb} for the normalization of all (anti)neutrino-nucleus cross-section measurements. Another pure-leptonic process, inverse muon decay $\nu_\mu e^- \to  \nu_e \mu^-$ and $\bar{\nu}_e e^- \to \bar{{\nu}}_\mu \mu^-$, requires (anti)neutrinos to be sufficiently energetic to produce the massive muon in the final state. The incoming energy should be larger than $10.9~\mathrm{GeV}$, which is slightly above the main region of modern artificial (anti)neutrino fluxes. 
Such high-energy tails are a very uncertain part of the (anti)neutrino beam~\cite{MINERvA:2016iqn} due to less-known hadroproduction cross sections for forward-going mesons in the direction of the proton beam. Recently, high-energy tails of the muon component in the incoming neutrino beam were also successfully constrained with the inverse muon decay (IMD) reaction $\nu_\mu e^- \to  \nu_e \mu^-$ by the MINERvA experiment~\cite{MINERvA:2021dhf}. According to the study in Ref.~\cite{Marshall:2019vdy}, the future DUNE experiment~\cite{DUNE:2020ypp} will have tens of thousands of elastic neutrino-electron scattering events and more than a few thousand inverse muon decay events. Consequently, both reactions will be accessed at the percent level, and radiative corrections become crucial for the correct interpretation of experimental measurements~\cite{Tomalak:2022jhh,Tomalak:2022xup}.

A comprehensive theoretical study of radiative corrections and various final-state distributions in elastic neutrino-electron scattering with error analysis was recently presented in Ref.~\cite{Tomalak:2019ibg} and compared to all previous calculations discussed in Refs.~\cite{Ram:1967zza,Weinberg:1967tq,tHooft:1971ucy,Sarantakos:1982bp,Bahcall:1995mm,Bardin:1983yb,Bardin:1985fg,Passera:2000ug,Green:1980uc,Bardin:1983zm,Zhizhin:1975kv,Byers:1979af,Salomonson:1974ys,Green:1980bd,Marciano:1980pb,Aoki:1980ix,Aoki:1981kq,Hioki:1981gi,Marciano:2003eq,Sirlin:2012mh}. Radiative corrections to the inverse muon decay were discussed in the ultrarelativistic limit in Ref.~\cite{Sarantakos:1982bp} and subsequently evaluated in Ref.~\cite{Bardin:1986dk}. Before the above-mentioned measurements by the MINERvA Collaboration, IMD results from the CHARM-II Collaboration~\cite{CHARM-II:1990xid,CHARM-II:1995xfh} have confirmed predictions of the Standard Model of particle physics.

The most relevant experimental observable is the muon energy spectrum with or without restrictions on the energy of the radiated real photon. The muon is scattered primarily in the forward direction. Although experimental resolution does not allow a precise  determination of the muon scattering angle, the muon angular distribution can potentially provide better selection criteria for IMD events. We compare our results for the muon energy spectrum with the Bardin-Dokuchaeva calculation for the dominant $\nu_\mu e^- \to \nu_e \mu^-$ channel~\cite{Bardin:1986dk} and provide a new result for the subdominant $\bar{\nu}_e e^- \to \bar{\nu}_\mu \mu^-$ process. We also provide new expressions for double-differential distributions in muon energy and muon scattering angle, in photon energy and photon scattering angle, as well as in muon energy and photon energy. We discuss how radiative corrections modify the experimentally sensitive distributions from Ref.~\cite{MINERvA:2021dhf}.

The paper is organized as follows. In Section~\ref{sec2}, we discuss the IMD reaction at tree level. We provide details of virtual radiative corrections in Appendix~\ref{sec3} and describe the evaluation of real contributions in Appendix~\ref{sec4}. We combine these calculations to obtain the resulting muon-energy spectrum at $\mathrm{O} \left( \alpha \right)$ precision in Section~\ref{sec5}, where we also present the total cross section. In the following Section~\ref{sec6}, we discuss distortions of experimentally accessed distributions due to $\mathrm{O} \left( \alpha \right)$ radiative corrections. We finish with conclusions and outlook in Section~\ref{sec7}. We provide new expressions for the triple-differential distribution in muon energy, muon scattering angle and photon energy, the double-differential distribution in muon energy and muon scattering angle, the double-differential distribution in photon energy and photon scattering angle, and the photon energy spectrum in the Supplemental material and Appendixes~\ref{app:3xsec}, \ref{app:2xsec_muon_energy_muon_angle}, \ref{app:2xsec_photon_energy_photon_angle}, and \ref{app:2xsec_photon_energy}, respectively.

\section{Inverse muon decay at tree level}
\label{sec2}

Consider muon production on atomic electrons by a neutrino beam, $ \nu_\mu \left( k_{\nu_\mu} \right) e^- \left( p_e \right) \to  {\nu}_e \left( k_{{\nu}_e} \right) \mu^- \left( p_\mu \right)$ [or $\bar{\nu}_e \left( k_{\bar{\nu}_e} \right) e^- \left( p_e \right) \to \bar{{\nu}}_\mu \left( k_{\bar{\nu}_\mu} \right) \mu^- \left( p_\mu \right)$].
This process is governed by the low-energy effective four-fermion interaction with scale-independent Fermi coupling constant $\mathrm{G}_\mathrm{F}$~\cite{Fermi:1934hr,Feynman:1958ty,Arason:1991ic,Antonelli:1980zt,Hill:2019xqk,Tomalak:2021lif}
\begin{align}
    {\cal L}_{\rm eff} = - 2 \sqrt{2} \mathrm{G}_\mathrm{F} 
  \bar{\nu}_{e}\gamma^\lambda \mathrm{P}_\mathrm{L} \nu_{\mu}
  \, \bar{\mu} \gamma_\lambda \mathrm{P}_\mathrm{L}  e + \mathrm{h.c.}. \label{eq:Lagrangian}
\end{align}
The reaction is kinematically allowed only for sufficiently high energies of the incoming neutrino $E_{\nu_\mu}, E_{\bar{\nu}_e}  \ge E^\mathrm{thr}_{\nu}$, where 
\begin{align}
    E^\mathrm{thr}_{\nu} = \frac{ m^2_\mu - m^2_e}{2 m_e}.
\end{align}
In radiation-free kinematics, the muon goes predominantly in the forward direction with a scattering angle $\theta_\mu$:
\begin{align}
    \cos \theta_{\mu} = \frac{ 2 \left( E_\nu E_\mu + m_e E_\mu - m_e E_\nu \right) - m^2_\mu - m^2_e}{2 E_\nu \sqrt{E^2_\mu - m^2_\mu}}.
\end{align}
The corresponding cone size increases with incoming (anti)neutrino energy. For example, the scattering within $0.2^\circ$ is allowed only for the incoming (anti)neutrino energy $E_\nu \gtrsim 38~\mathrm{GeV}$.

The differential cross section with respect to the muon energy $E_\mu$, as a function of the incoming neutrino energy $E_{\nu}$, is given by 
\begin{align}
\frac{\mathrm{d} \sigma_\mathrm{LO} \left( \nu_\mu e^- \to {\nu}_e \mu^- ,~\bar{\nu}_e e^- \to \bar{{\nu}}_\mu  \mu^- \right)}{\mathrm{d} E_{\mu}} = \frac{|\mathrm{T}_\mathrm{LO}|^2 }{32 \pi m_e E^2_{\nu}} \,, \label{eq:LO_differential}
\end{align}
where the squared matrix element at leading order is  $|\mathrm{T}_\mathrm{LO}|^2 =  64 \mathrm{G}_\mathrm{F}^2  p_\mu \cdot k_{{\nu}_e} p_e \cdot k_{\nu_\mu}$ 
for 
$ \nu_\mu \left( k_{\nu_\mu} \right) e^- \left( p_e \right) \to  {\nu}_e \left( k_{{\nu}_e} \right) \mu^- \left( p_\mu \right)$ 
and  $|\mathrm{T}_\mathrm{LO}|^2 =  64 \mathrm{G}_\mathrm{F}^2  p_\mu \cdot k_{\bar{\nu}_e}  p_e \cdot k_{\bar{\nu}_\mu}$ 
for $\bar{\nu}_e \left( k_{\bar{\nu}_e} \right) e^- \left( p_e \right) \to \bar{{\nu}}_\mu \left( k_{\bar{\nu}_\mu} \right) \mu^- \left( p_\mu \right)$. Integration of this distribution over the kinematically allowed range of muon energies, 
\begin{align}
E_\mu^\mathrm{min} = \frac{m^2_\mu + m^2_e}{2 m_e} \le E_\mu \le \frac{\left( E_{\nu} + \frac{m_e}{2} \right)^2 + \frac{m^2_\mu}{4}}{E_{\nu} + \frac{m_e}{2}} = E_\mu^\mathrm{max} \,, \label{eq:muon_energy_range}
\end{align}
results in the following total cross sections~\cite{Bardin:1986dk}:
\begin{align}
\sigma_\mathrm{LO} \left( \nu_\mu e^- \to {\nu}_e \mu^- \right) &= \frac{2 \mathrm{G}_\mathrm{F}^2 m_e \left( E_\mu^\mathrm{max} - E_\mu^\mathrm{min} \right)}{\pi} \frac{  E_{\nu_\mu} - E^\mathrm{thr}_{\nu}}{E_{\nu_\mu} }, \label{eq:tree_level_inverse_muon_decay_xsection_neutrino} \\
\sigma_\mathrm{LO}^{\bar{\nu}} \left( \bar{\nu}_e e^- \to \bar{{\nu}}_\mu \mu^- \right) &= \frac{2 \mathrm{G}_\mathrm{F}^2 m_e \left( E_\mu^\mathrm{max} - E_\mu^\mathrm{min} \right)}{\pi} \left[ \frac{   \left(  E_\mu^\mathrm{max}\right)^2 + E_\mu^\mathrm{max} E_\mu^\mathrm{min} + \left(  E_\mu^\mathrm{min}\right)^2}{3 E^2_{\bar{\nu}_e} } + \frac{  E_{\bar{\nu}_e}+m_e}{E_{\bar{\nu}_e} } \frac{  E_{\bar{\nu}_e} + E_\mu^\mathrm{min}}{E_{\bar{\nu}_e} }  \right.\nonumber \\
&\left. - \frac{  E_{\bar{\nu}_e} + E_\mu^\mathrm{min}  -\frac{E^\mathrm{thr}_{\nu}}{2} }{E_{\bar{\nu}_e} } \frac{     E_\mu^\mathrm{max} +   E_\mu^\mathrm{min}}{E_{\bar{\nu}_e} } \right]. \label{eq:tree_level_inverse_muon_decay_xsection_antineutrino} 
\end{align}

We provide details of the standard calculation of real and virtual radiative corrections to the inverse muon decay cross sections in Appendixes~\ref{sec3} and \ref{sec4}, respectively.

\section{Muon energy spectrum and integrated cross section}
\label{sec5}

Adding virtual and real corrections, we obtain 
the muon energy spectrum for inverse muon decay, $\nu_\mu e^- \to {\nu}_e \mu^-$, including photons of arbitrarily large energy allowed by kinematics.%
\footnote{Note that this inclusive observable is independent of the infrared regulators $\lambda$ and $\Delta E$ from Appendixes~\ref{sec3} and \ref{sec4}.
}
In the limit $E_\nu \gg m_e$ (i.e., neglecting order $m_e/E_\nu$ power corrections) our results are in agreement with the calculation of Bardin and Dokuchaeva~\cite{Bardin:1986dk}:
\begin{align}
\frac{\mathrm{d} \sigma}{\mathrm{d} E_\mu } &= \frac{\mathrm{d} \sigma_\mathrm{LO}}{\mathrm{d} E_\mu } + \frac{2 \mathrm{G}_\mathrm{F}^2 m_e}{\pi} \frac{\alpha}{\pi} \left[ \frac{1+3 x}{2} \left( \mathrm{Li}_2 \frac{1-\frac{x}{y}}{1-x} -  \mathrm{Li}_2 \frac{y-x}{1-x} - \ln \frac{y}{x} \ln \frac{y-x}{1-x} \right) \right. \nonumber \\
&\left. + \left( 1 -x \right) \left( \left( \ln \frac{y^2}{x r_e} - 2 \right) \ln \frac{y - x}{y}  + \ln \frac{y}{x} \ln \left( 1 - y \right) - \mathrm{Li}_2 x + \mathrm{Li}_2 y + \mathrm{Li}_2 \frac{x-y}{1-y} + \frac{3}{2} \left( 1 -x \right) \ln \left( 1 - x \right) \right) \right. \nonumber \\
&\left. - \frac{7 x^3}{36} y^{-3} + \frac{x^2}{12} \left( 1 + \frac{7 x}{2} \right) y^{-2} + \left( - \frac{7 x}{12} - \frac{x^2}{2} - \frac{x^3}{6} \right) y^{-1}  -\frac{47}{36} + \frac{25 x}{8} + \frac{3x^2}{8} - \left( \frac{11}{12} + \frac{x}{4} \right) y + \frac{y^2}{24} \right. \nonumber \\
&\left. -  \left(\frac{x^2}{2} y^{-2} + \left( \frac{x}{2} - 2 x^2 \right) y^{-1} + \frac{1}{4} - \frac{3x}{4} + \frac{3 x^2}{2}+ \frac{y}{2} \right) \ln x + \left( x^2 y^{-2} + x \left( 1 - 4 x \right) y^{-1} + \frac{3x^2}{2} + y \right) \ln y \right. \nonumber \\
&\left. +  \left( \frac{x^3}{6} y^{-3} - \frac{x^2 \left(1 + x \right)}{4} y^{-2} + \frac{x \left( 1+3 x \right)}{2} y^{-1} - \frac{23}{12} + \frac{9x}{4} - \frac{3 x^2}{2} - \frac{y}{2} \right) \ln \left( 1 - y \right) \right. \nonumber \\
&\left. + \left( \frac{x^2}{6} y^{-2} - \frac{x}{4} \left( \frac{1}{3} +x \right) y^{-1} + \frac{5}{4} \left( \frac{1}{3} +x \right) + \frac{y}{2} \right) \frac{y-x}{y}\ln \frac{y-x}{y} \right. \nonumber \\
&\left. -  \left( \frac{x^3}{6} y^{-3} + \frac{x^2 \left( 1 - x \right)}{4} y^{-2} + \left( x - \frac{x^2}{2} \right)y^{-1} - \frac{2}{3} \right) \ln r_e \right],\label{eq:muon_energy_spectrum} \\
x&= \frac{m^2_\mu}{2m_e E_{\nu_\mu}}, \qquad \qquad \qquad  y = \frac{E_\mu}{E_{\nu_\mu}}, \qquad \qquad \qquad r_e = \frac{m_e}{2 E_{\nu_\mu}}. \label{eq:muon_energy_spectrum_notations}
\end{align}
Comparing the double-differential distribution in photon energy and muon energy in Ref.~\cite{Bardin:1986dk} to our numerical evaluation, which starts from the matrix element, Eq.~(\ref{eq:matrix_element_radiative_inverse_muon_decay}) in Appendix~\ref{sec4}, and performs numerical integration within the allowed phase space of the process, we find agreement only inside the range of photon energies $ \left( m_e -  \left(  E_\mu - \sqrt{E_\mu^2 - m^2_\mu} \right)  \right)/2 \le k_\gamma \le \left( m_e + 2 E_\nu - \left( E_\mu + \sqrt{E_\mu^2 - m^2_\mu} \right) \right) /2 $; in particular, the double-differential distribution in the calculation of Bardin and Dokuchaeva is not positive-definite in small end-point regions outside this range.%
\footnote{Our result agrees up to power corrections in the electron mass with the expression of Ref.~\cite{Bardin:1986dk} [their Eq. (24)] for the contribution to the muon energy spectrum from events with photons above an energy cutoff $\Delta E$ when $\Delta E \gg m_e$ and $\left( m_e + 2 E_\nu - \left( E_\mu + \sqrt{E_\mu^2 - m^2_\mu} \right) \right) /2 - \Delta E  \gg m_e $.}

The corresponding muon energy spectrum for $\bar{\nu}_e e^- \to  \bar{\nu}_\mu \mu^-$ in the limit $E_\nu \gg m_e$ is given by
\begin{align}
\frac{\mathrm{d} \sigma^{\bar{\nu}}}{\mathrm{d} E_\mu } &= \frac{\mathrm{d} \sigma^{\bar{\nu}}_\mathrm{LO}}{\mathrm{d} E_\mu } + \frac{2 \mathrm{G}_\mathrm{F}^2 m_e}{\pi} \frac{\alpha}{\pi} \left( 1 - y \right) \left[ \left(1+x -y \right)\left( \mathrm{Li}_2 \frac{1-\frac{y}{x}}{1-y} +  \mathrm{Li}_2 \frac{y-x}{y} + \ln \frac{x-y}{y} \ln \frac{y^2}{x r_e} + \frac{1}{2} \ln^2 \frac{y}{x}  \right) \right. \nonumber \\
&\left. + \frac{1 - 2 \left( 1 - y \right) y + x \left( 1 + y \right)}{ 2\left( 1 -y \right)} \left( \mathrm{Li}_2 \frac{y-1}{y} + \mathrm{Li}_2 \frac{x-y}{x} - \mathrm{Li}_2 \frac{x-1}{x}+ \ln \frac{y}{x} \ln \left[ y \left( y - x \right) \right] - \ln y \ln \left( 1 - y \right)  \right) \right. \nonumber \\
&\left. + \frac{1 - 2 \left( 1 - y \right) y + x \left( 1 + y \right)}{ 2\left( 1 -y \right)} \ln x \ln \left( 1 - x \right) + \frac{x^3}{18} y^{-3} - \frac{x^2}{24} \left( 7 - \frac{5 x}{3} \right) y^{-2} - \left( \frac{4 x}{3} + \frac{23x^2}{24} \right)  y^{-1}  -\frac{31}{72}  \right. \nonumber \\
&\left.+ \frac{5 x}{24} + \frac{49}{72}y -  \left(\frac{x^2}{2} y^{-2} + \frac{x \left( 1 - x \right)}{2} y^{-1} - \frac{1}{2} - \frac{9x}{4} \right) \ln x + \frac{3}{4} \frac{\left( 1 - x \right) \left( 1 - x - 2 y \right) }{1-y} \ln \left( 1 - x \right)\right. \nonumber \\
&\left. + \left( x^2 y^{-2} + x \left( 1 - 2 x \right) y^{-1}  -\frac{3}{4} - \frac{5}{2} x + \frac{x^2}{4} + \left( \frac{1}{2} + \frac{3}{2} x \right) y + y^2 \right) \frac{\ln y}{1-y}  \right. \nonumber \\
&\left. +  \left( \frac{x^3}{6} y^{-3} - \frac{x^2 \left(3 - x \right)}{12} y^{-2} + \frac{x \left( 2+3 x \right)}{4} y^{-1} - \frac{7}{6} - \frac{x}{4} + \frac{5}{3} y \right) \ln \left( 1 - y \right) \right. \nonumber \\
&\left. - \left( \frac{x^3}{6} y^{-3} - \frac{x^2 \left( 3 + x \right)}{12} y^{-2} + \frac{x}{2} \left( 1 + 2 x - \frac{x^2}{6} \right) y^{-1}+\frac{19}{12} - \frac{x}{4} - \left( \frac{8}{3} + \frac{x}{4}\right) y + \frac{y^2}{3} \right) \frac{1}{1-y}\ln \frac{y-x}{y} \right. \nonumber \\
&\left. -  \left( \frac{x^3}{6} y^{-3} + \frac{x^2 \left( 3 + x \right)}{12} y^{-2} + \left( x + \frac{x^2}{4} \right)y^{-1} - \frac{2}{3} - x + \frac{2}{3} y \right) \ln r_e \right], \label{eq:muon_energy_spectrum_antineutrino}
\end{align}
where $x,y$,~and~$r_e$ are given by the substitution $E_{\nu_\mu} \to E_{\bar{\nu}_e}$ in Eqs.~(\ref{eq:muon_energy_spectrum_notations}).

\begin{figure}[tp]
\centering
\includegraphics[width=0.79\textwidth]{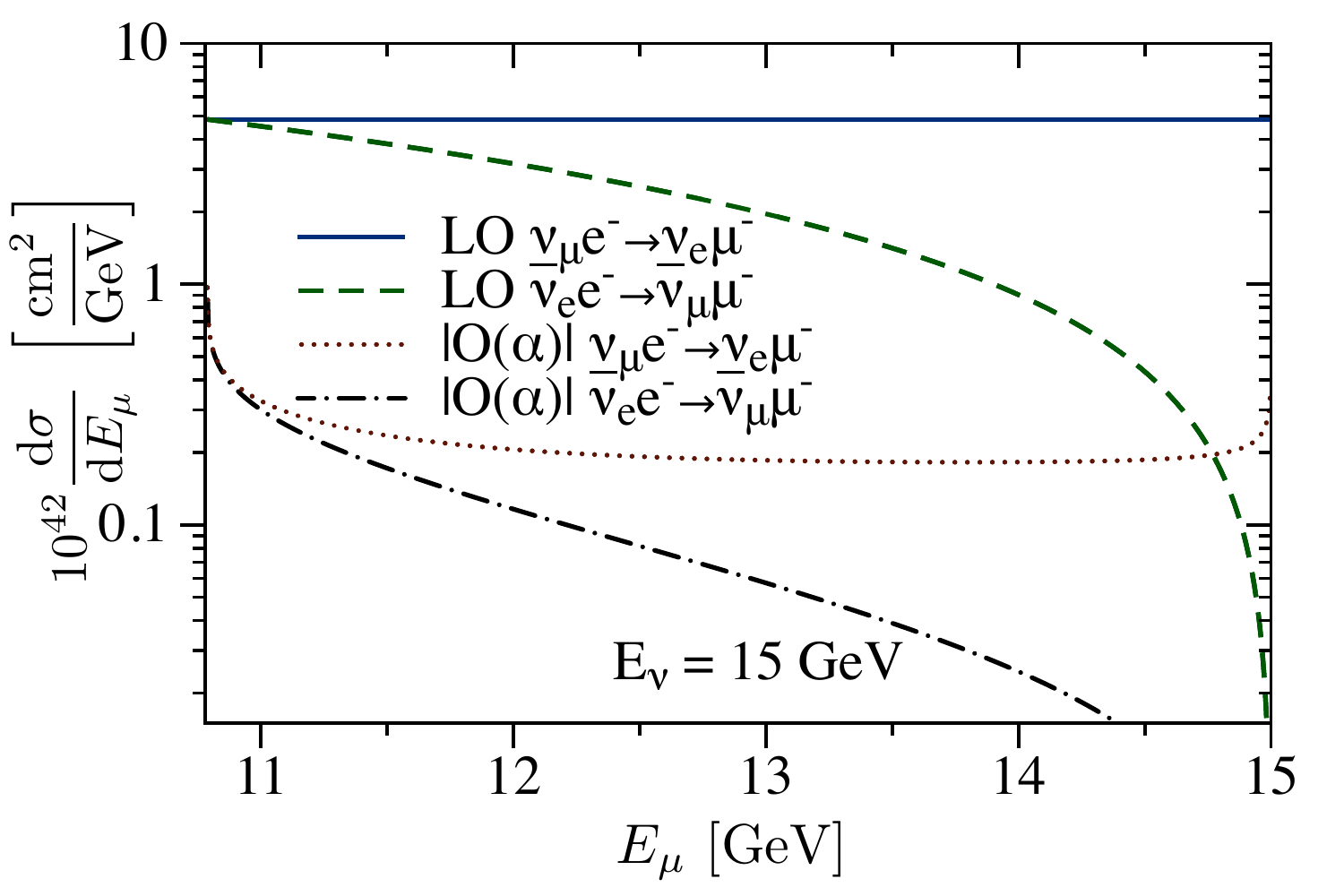}
\caption{Leading-order muon spectrum and $\mathrm{O} \left( \alpha \right)$ corrections for a fixed neutrino energy $E_\nu = 15$~GeV. The $\nu_\mu e^- \to {\nu}_e \mu^-$ process at leading order is shown by the blue solid line and is compared to the spectrum in $\bar{\nu}_e e^- \to \bar{{\nu}}_\mu \mu^-$ at leading order, which is shown by green dashed line, and to the $\mathrm{O} \left( \alpha \right)$ contribution in Eqs.~(\ref{eq:muon_energy_spectrum}) and~(\ref{eq:muon_energy_spectrum_antineutrino}), cf. the red dotted and black dash-dotted lines respectively. The $\mathrm{O} \left( \alpha \right)$ contribution is negative, i.e., decreases the total and differential cross sections at all values of muon energy.\label{fig:Emu_spectrum_15GeV}}
\end{figure}

At the fixed illustrative neutrino energy of $E_\nu =15$~GeV, Fig.~\ref{fig:Emu_spectrum_15GeV} shows muon energy spectra for the tree-level processes $\nu_\mu e^- \to {\nu}_e \mu^-$ and $\bar{\nu}_e e^- \to \bar{{\nu}}_\mu \mu^-$ as well as the $\mathrm{O} \left( \alpha \right)$ contribution to $\nu_\mu e^- \to {\nu}_e \mu^-$ from Eq.~(\ref{eq:muon_energy_spectrum}) and $\bar{\nu}_e e^- \to \bar{{\nu}}_\mu \mu^-$ from Eq.~(\ref{eq:muon_energy_spectrum_antineutrino}). We show the latter with an opposite sign for convenience. The radiative corrections reduce the cross section by $3$-$4~\%$. They have the largest relative size for backward scattering and increase going to forward angles.

Integrating the muon energy spectrum over the kinematically allowed range in Eq.~(\ref{eq:muon_energy_range}), we obtain the $\mathrm{O} \left( \alpha \right)$ contribution to the unpolarized inverse muon decay cross section $\sigma$. For illustration, we present two limits of interest. The leading term in $m_e/E_\nu$ expansion is given by
\begin{align}
\sigma & \underset{r_e \ll 1}{\longrightarrow} \sigma_\mathrm{LO} + \frac{2 \mathrm{G}_\mathrm{F}^2 m_e E_\nu}{\pi} \frac{\alpha}{\pi} \left[ \frac{1}{24} \left(19 - 4 \pi^2 \left( 1 - \frac{3}{2} x + \frac{5}{2} x^2 \right) + 16 \ln r_e + 36 x \left(1 - 2 \ln x \right) + x^2 \left( 45 - 4 x \right) \right) \right.\nonumber \\ 
&\left. + \frac{x}{2} \left( \left( 1 + 3 x \right) \mathrm{Li}_2 x + \left( 3 - 7 x \right) \left( \mathrm{Li}_2 \left( 1 - \frac{1}{x}\right) + \frac{1}{2} \ln^2 x \right)  - 8 + x \ln x \ln r_e \right) - x \ln x \left( 1 - \frac{13}{4} x\right) \right.\nonumber \\ 
&\left. - \ln \left( 1 - x \right) \left( \left( 1 - \frac{1}{2} x - \frac{5}{2} x^2 \right) \ln x + \left( 1 - x \right)^2 \left( 4 + \ln r_e \right) \right)  - x \left( 1 - \frac{x\left(3-x \right)}{6} \right)\ln r_e\right], \\
\sigma^{\bar{\nu}} & \underset{r_e \ll 1}{\longrightarrow} \sigma^{\bar{\nu}}_\mathrm{LO} + \frac{2 \mathrm{G}_\mathrm{F}^2 m_e E_\nu}{\pi} \frac{\alpha}{\pi} \left[ \frac{1}{72} \left(43 - 4 \pi^2 \left( 1 + 9 x + \frac{3}{2} x^2 - x^3 \right) + 16 \ln r_e + \frac{27 x}{2}\left(9 - 4 \ln x \right) - 41 x^2\right)  \right.\nonumber \\ 
&\left. + \frac{25 x^3}{144}   + \frac{x}{2} \left( \left( 7 + x - x^2 \right) \mathrm{Li}_2 x - \left( 3 + x - \frac{x^2}{3} \right) \left( \mathrm{Li}_2 \left( 1 - \frac{1}{x}\right) + \frac{1}{2} \ln^2 x \right) - \frac{34}{9} + \frac{x^2}{2} \ln x \ln r_e \right) \right.\nonumber \\ 
&\left.- \frac{x}{3} \ln x \left( 1 + \frac{5}{4} x - \frac{47}{24} x^2 \right)   - \ln \left( 1 - x \right) \left( \left( 1 - 6 x - \frac{3}{2} x^2 + x^3 \right) \frac{\ln x}{3} + \left( 1 - \frac{3x}{2} + \frac{x^3}{2} \right) \frac{\ln r_e}{3}  \right) \right.\nonumber \\ 
&\left. - \left( \frac{17}{9} + \frac{x \left( 7 - 11 x \right)}{18} \right) \left( 1 - x \right) \ln \left( 1 - x \right) - \frac{x}{3} \left( 1 - \frac{x\left(3+x \right)}{12} \right)\ln r_e\right],
\end{align}
while the high-energy limit, $x \ll 1$, is given by
\begin{align}
\sigma \left( E_\nu \right) - \sigma_\mathrm{LO} \left( E_\nu \right)& \underset{x \ll 1}{\longrightarrow}  \frac{\mathrm{G}_\mathrm{F}^2 m_e E_\nu}{12\pi} \frac{\alpha}{\pi} \left( 19 - 4 \pi^2 + 16 \ln r_e + 36 x \left(1 - 2 \ln x \right) \right), \\
\sigma^{\bar{\nu}} \left( E_\nu \right) - \sigma^{\bar{\nu}}_\mathrm{LO} \left( E_\nu \right)& \underset{x \ll 1}{\longrightarrow}  \frac{\mathrm{G}_\mathrm{F}^2 m_e E_\nu}{36\pi} \frac{\alpha}{\pi} \left( 43 - 4 \pi^2 \left( 1 + \frac{9}{2} x \right) + 16 \ln r_e+ \frac{27}{2} x \left( 9 - 4 \ln x \right) \right),
\end{align}
where the leading terms at $x\to 0$ coincide with the well-known expressions in Ref.~\cite{Sarantakos:1982bp}.

We provide the total cross section, double-differential distribution in muon energy and muon scattering angle, double-differential distribution in muon energy, and triple-differential distribution in muon energy, muon scattering angle, and photon energy in  Appendixes~\ref{app:3xsec}, \ref{app:2xsec_muon_energy_muon_angle}, and in the Supplemental material.

\section{Distortion of experimentally accessed distributions}
\label{sec6}

Experimentally, inverse muon decay events are distinguished from other reactions by looking for high-energy muons, above the $E_\mu^\mathrm{min}$ of Eq.~(\ref{eq:muon_energy_range}) with no other particles in the final state, and which are along the direction of the incoming neutrino due to the kinematics of elastic scattering from electrons.  Radiative corrections cause events with real photons in the final state and with a different distribution of muon energies and angles than in the tree-level process.  This section explores those changes from the tree-level predictions.

Experiments will need to reject events from $\nu_\mu$ quasielastic scattering on nucleons in nuclei which may appear to be consistent with elastic kinematics, but which will have a recoiling proton in the final state.  Similarly, inelastic processes can produce high-energy forward muons with other particles in the final state.  Because there are many possible elastic and inelastic reactions, a common experimental strategy is to remove events with any other visible energy than the muon in the final state. An energetic real photon from radiative processes, even one nearly collinear with the muon, may produce visible energy that will veto the event due to this requirement.

\begin{figure}[tp]
\centering
\includegraphics[width=0.79\textwidth]{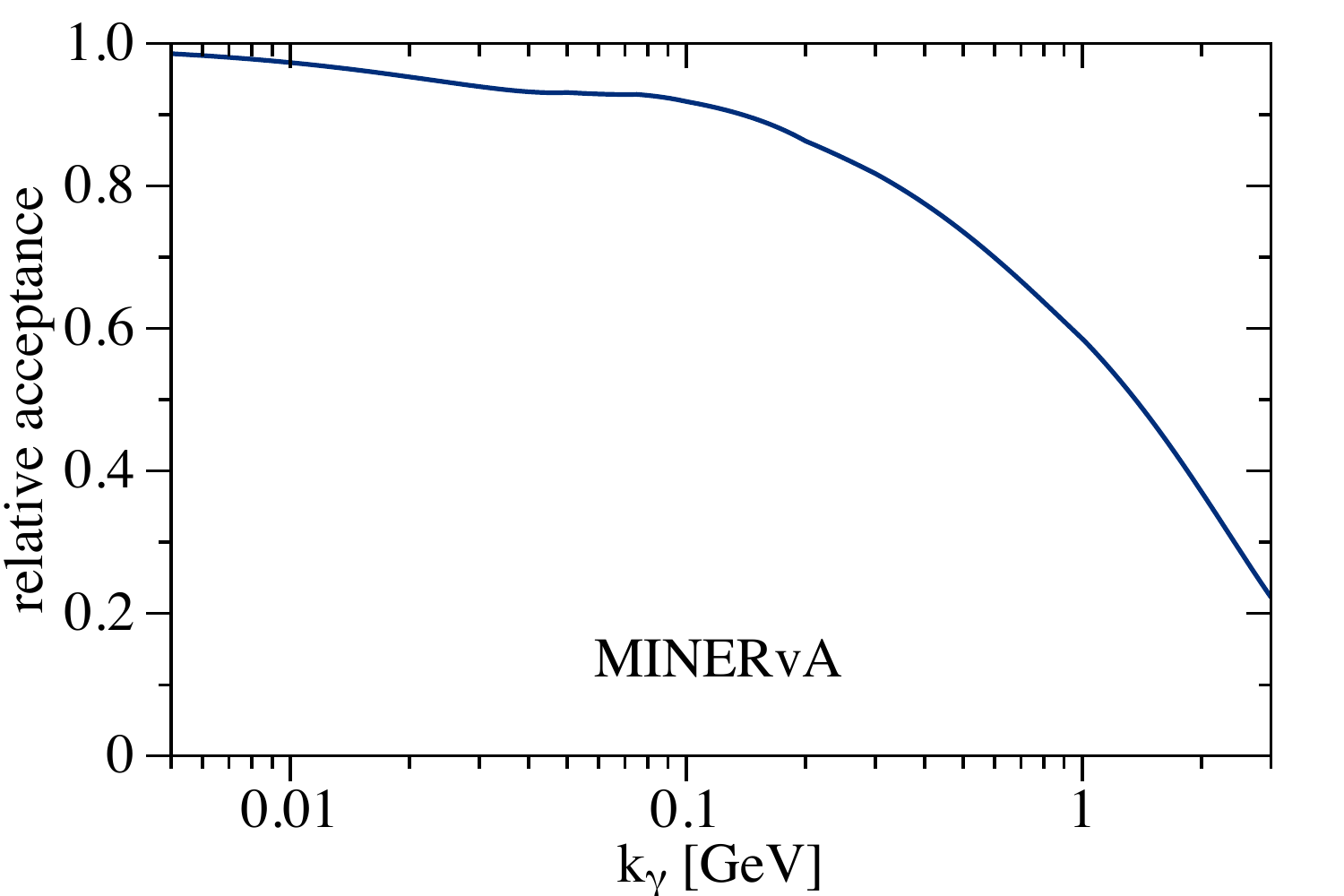}
\caption{MINERvA's probability to accept a radiative IMD event as a function of collinear photon energy~\cite{MINERvA:2021dhf}. \label{fig:veto_probability}}
\end{figure}

While this experimental strategy will be common to all measurements, the details of the effect will be particular to each experimental setup. In its analysis~\cite{MINERvA:2021dhf}, MINERvA predicted the relative acceptance as a function of photon energy, and that prediction is shown in Fig.~\ref{fig:veto_probability}.

\begin{figure}[tp]
\centering
\includegraphics[width=1.00\textwidth]{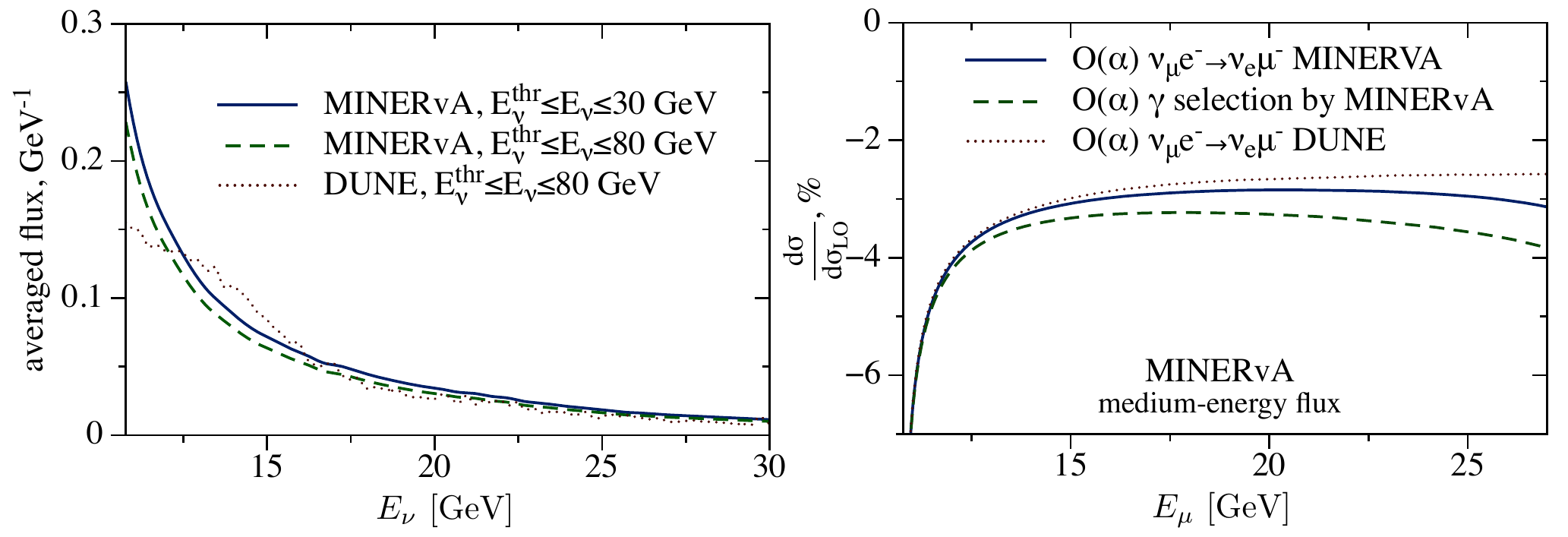}
\caption{Left panel: MINERvA medium-energy flux is averaged over the energies above the IMD threshold to $30$ and $80~\mathrm{GeV}$. For DUNE experiment, we present the averaging over the range up to $80~\mathrm{GeV}$. Right panel: Ratio of $\mathrm{O} \left( \alpha \right)$ contribution to the leading-order result for the muon energy spectrum above $E_\mu^\mathrm{min}$, averaged over the anticipated DUNE flux, shown by the red dashed line, is compared to this ratio averaged over the medium-energy flux of the MINERvA experiment, shown by the blue solid line. The green dashed line shows the further reduction in cross section due to the probability of vetoing the event due to the presence of a real radiated photon in MINERvA's analysis of IMD events~\cite{MINERvA:2021dhf}. MINERvA's probability to accept events with real photons as a function of the photon energy is shown in Figure~\ref{fig:veto_probability}.\label{fig:flux_averaged}}
\end{figure}

Averaging over high-energy tails of the expected flux in the DUNE experiment~\cite{dune_page} and medium-energy ``neutrino'' (forward horn current) mode for the MINERvA experiment~\cite{MINERvA:2016iqn,MINERvA:2016ing,MINERvA:2019hhc}, we provide the effect of $\mathrm{O} \left( \alpha \right)$ on muon energy spectra for two representative examples of neutrino experiments that do or will use IMD to constrain its high-energy flux tails in Fig.~\ref{fig:flux_averaged}. For MINERvA and DUNE predictions, we average over the (anti)neutrino energy above the threshold value $E_\nu^\mathrm{thr}$ but below $30$ and $80~\mathrm{GeV}$ respectively.  We illustrate this averaging in the left panel of Fig.~\ref{fig:flux_averaged} and compare it to fluxes averaged over the same region in both experiments. The average over the flux decreases the resulting cross section compared to the fixed energy $E_\nu = 15~\mathrm{GeV}$ result, which is shown in Fig.~\ref{fig:Emu_spectrum_15GeV}, since the flux falls monotonically with (anti)neutrino energy and is convoluted with slower rising cross section. Distortions of the muon energy spectrum increase as the neutrino energy approaches the threshold of the inverse muon decay from above.  

The effect on the measurable cross section from the removal of events with real photons by MINERvA is also shown in Fig.~\ref{fig:flux_averaged} and compared with the $\mathrm{O} \left( \alpha \right)$ correction. It is less than a $1\%$ reduction in the observed rate, with a larger effect for higher muon energies.

The kinematics of elastic scattering from electrons produces a relationship between the muon energy and angle with respect to the incoming neutrino direction. A useful combination is 
\begin{align}
{\cal{F}} \left( E_\mu,~\theta_\mu \right) \equiv E_\mu \theta_\mu^2  \approx \left( 1 - \frac{E_\mu}{E_\nu}\right) \left( 2 m_e - \frac{m_\mu^2}{E_\mu} \right). \label{eq:variable_F_DUNE}
\end{align}
When $E_\nu \gg E_\mu$ and $E_\mu \gg E_\mu^\mathrm{min}$, ${\cal{F}}$ can approach its upper limit of $2m_e$.   

In measurements of elastic neutrino-electron scattering by the MINERvA experiment~\cite{MINERvA:2015nqi,MINERvA:2019hhc,MINERvA:2022vmb}, the same quantity was used to select events that were due to elastic scattering from electrons.  In this case $E_e \gg E_e^\mathrm{min}$ for all of the selected events.  However, for IMD for the experimental fluxes considered above from DUNE and MINERvA, neither condition above is true for most events, and therefore typically ${\cal{F}} \ll 2m_e$.  In particular because the factor $\left( 1-\frac{E_\mu}{E_\nu}\right) $ is usually small, one might want to consider an ``idealized" version of ${\cal{F}}$,
\begin{align}
{{\cal{F}}^{\rm\scriptstyle ideal}} \left( E_\mu,~\theta_\mu \right) \equiv \frac{E_\mu \theta_\mu^2}{ 1 - \frac{E_\mu}{E_\nu}}\approx  2 m_e - \frac{m_\mu^2}{E_\mu} . \label{eq:variable_F_HE}
\end{align}
However, this quantity is not accessible since the neutrino energy is not known on an event-by-event basis.

In the measurement by the MINERvA experiment~\cite{MINERvA:2021dhf}, the analysis enforced elastic kinematics for a ``maximum" energy of likely candidate events in its beam.  The variable ${\cal{F}}^{\rm\scriptstyle MINERvA}$,
\begin{align}
{{\cal{F}}^{\rm\scriptstyle MINERvA}} \left( E_\mu,~\theta_\mu \right) \equiv \frac{E_\mu \theta_\mu^2}{1 - \frac{E_\mu}{E^\mathrm{max}}}, \label{eq:variable_F}
\end{align}
with $E^\mathrm{max} = 35~\mathrm{GeV}$, was used for the selection of signal events by placing a cut on $ {{\cal{F}}^{\rm\scriptstyle MINERvA}} \left( E_\mu,~\theta_\mu \right)$.

\begin{figure}[tp]
\centering
\includegraphics[width=0.79\textwidth]{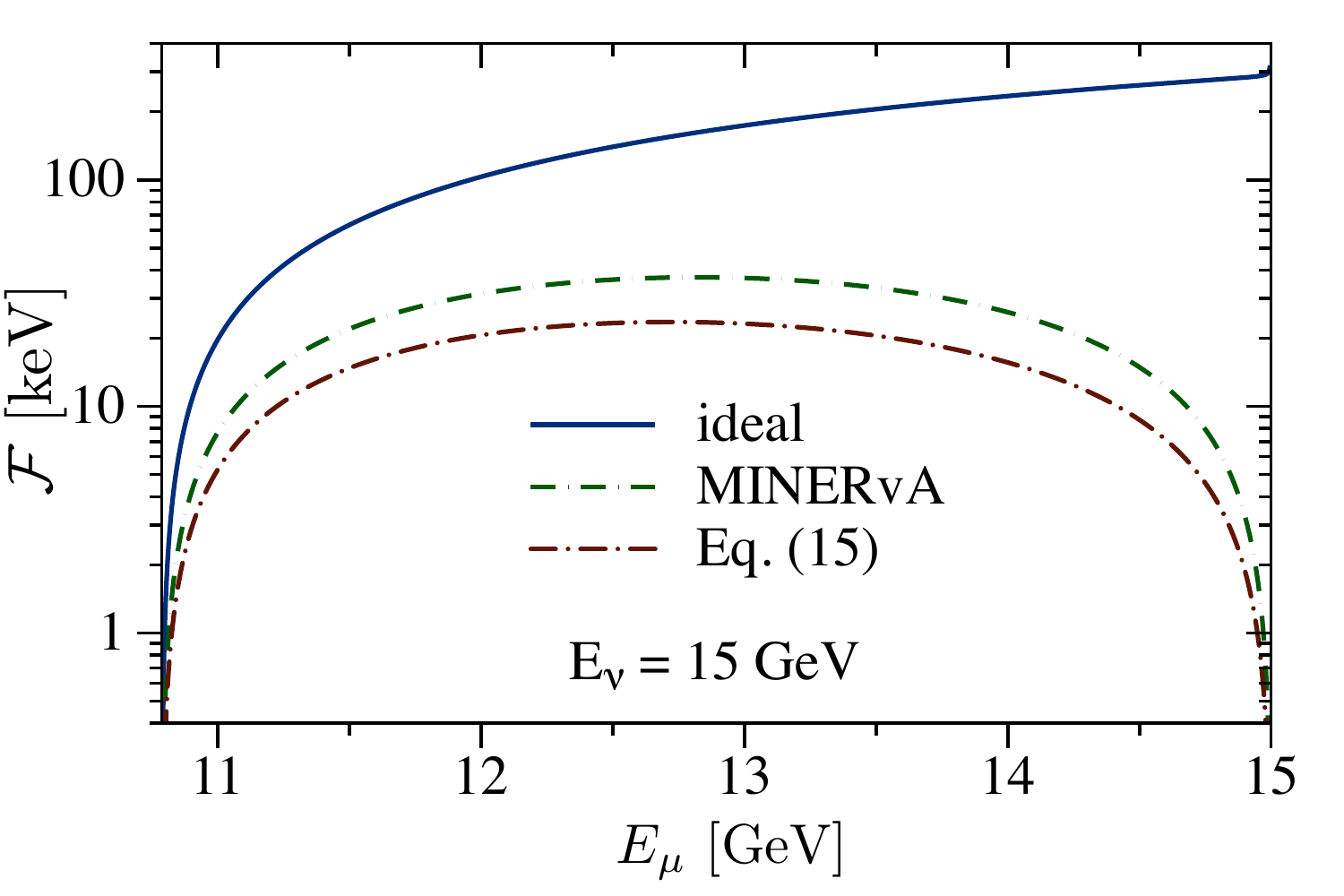}
\caption{The variable $\cal{F}$ as defined in Eqs.~(\ref{eq:variable_F_DUNE}),~(\ref{eq:variable_F_HE}),~and~(\ref{eq:variable_F}) is presented as a function of the muon energy $E_\mu$ at the fixed neutrino energy $E_\nu = 15$~GeV .\label{fig:differentF_15GeV}}
\end{figure}

\begin{figure}[tp]
\centering
\includegraphics[width=0.79\textwidth]{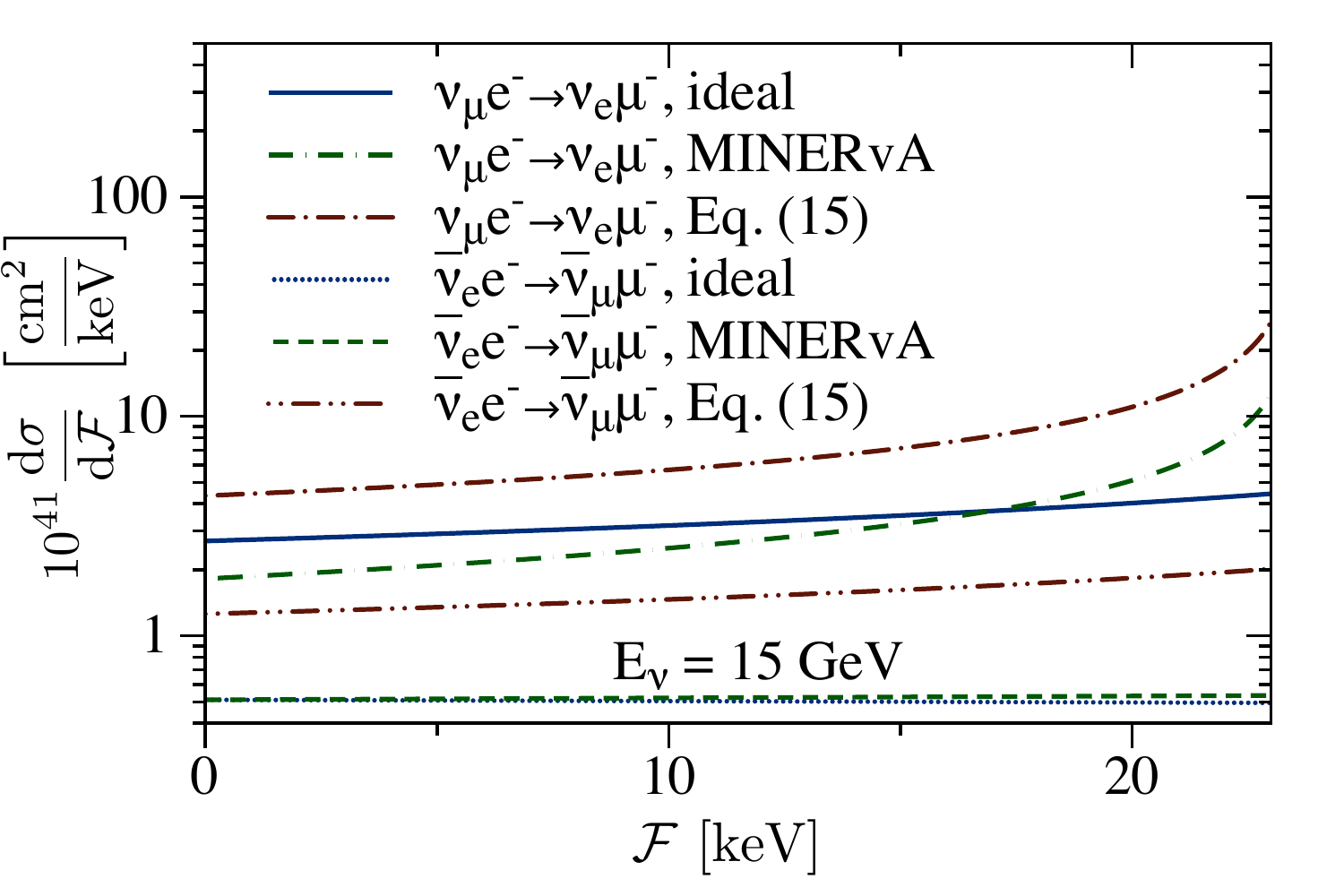}
\caption{Distribution of the variable $\cal{F}$ for the tree-level events in IMD reaction according to definitions in Eqs.~(\ref{eq:variable_F_DUNE}),~(\ref{eq:variable_F_HE}),~and~(\ref{eq:variable_F}) is shown at the fixed neutrino energy $E_\nu = 15$~GeV.\label{fig:differentF_spectrum_15GeV}}
\end{figure}

To illustrate various definitions for the variable $\cal{F}$, we present all three variants as a function of the final-state muon energy $E_\mu$ for the fixed neutrino energy $E_\nu = 15$~GeV in Fig.~\ref{fig:differentF_15GeV}. The size of this variable in the inverse muon decay is below $10-100~\mathrm{keV}$. $\cal{F}$ vanishes both in forward and backward directions for definitions in Eqs.~(\ref{eq:variable_F_DUNE}) and (\ref{eq:variable_F}) contrary to the forward scattering only for the definition in Eq.~(\ref{eq:variable_F_HE}). In Fig.~\ref{fig:differentF_spectrum_15GeV}, we also present the tree-level distributions of the variable $\cal{F}$ for the same illustrative neutrino energy, in the region that is allowed kinematically for all three definitions. We observe a significant redistribution of events moving from one definition of the variable $\cal{F}$ to another.

\begin{figure}[tp]
\centering
\includegraphics[width=0.79\textwidth]{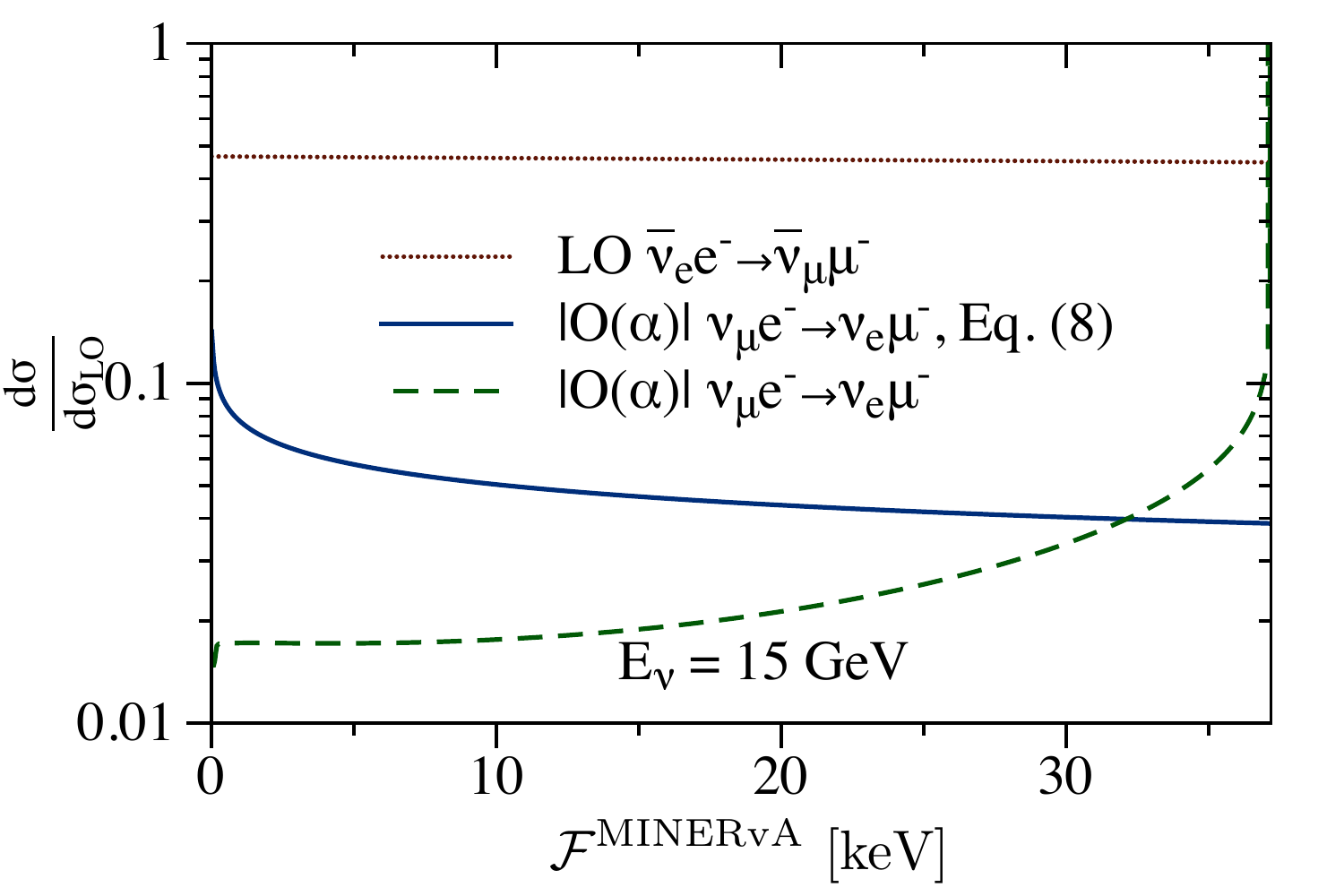}
\caption{Comparisons of leading order and $\mathrm{O} \left( \alpha \right)$ correction in the distribution of the variable ${\cal{F}}^\mathrm{MINERvA}$ for a fixed neutrino energy $E_\nu = 15$~GeV. The ratio of the leading-order processes, $\nu_\mu e^- \to {\nu}_e \mu^-$ to $\bar{\nu}_e e^- \to \bar{{\nu}}_\mu \mu^-$ is almost constant, as is shown by the red dotted line. The $\mathrm{O} \left( \alpha \right)$ correction from Eq.~(\ref{eq:muon_energy_spectrum}) is shown by the blue solid line under the assumption that the kinematics is identical to that of radiation-free scattering. It is compared to the ``true" $\mathrm{O} \left( \alpha \right)$ contribution, which is obtained by integrating the appropriate double-differential distribution and adding virtual and soft-photon corrections. Note that both $\mathrm{O} \left( \alpha \right)$ contributions are negative and so decrease the cross section.\label{fig:F_spectrum_15GeV}}
\end{figure}

To illustrate the effect of radiative corrections on the distribution of the ${\cal F}$ variables, we keep MINERvA's definition in Eq.~(\ref{eq:variable_F}) for applications to MINERvA's study~\cite{MINERvA:2016iqn,MINERvA:2016ing}. However for a general experiment, including the application to the DUNE flux~\cite{dune_page} in this paper, we do not wish to enforce a maximum neutrino energy above which we would drop the constraint, so we study instead the original ${\cal{F}}$ of Eq.~(\ref{eq:variable_F_DUNE}). We present in Fig.~\ref{fig:F_spectrum_15GeV} the distribution of the variable ${\cal{F}}^\mathrm{MINERvA}$ at tree level for the fixed incoming neutrino energy $E_\nu = 15~\mathrm{GeV}$, and compare it to the $\mathrm{O} \left( \alpha \right)$ contribution of radiative corrections by integrating the double-differential distribution in muon energy and muon scattering angle, and by providing the naive estimate assuming the kinematics of the radiation-free process and Eq.~(\ref{eq:muon_energy_spectrum}). $\mathrm{O} \left( \alpha \right)$ contributions shift the distribution of ${\cal{F}}^\mathrm{MINERvA}$ variable by a percent-level correction. Note also that all inverse muon decay events from neutrinos of energy $E_\nu \le 30~\mathrm{GeV}$ belong to the first bin in the variable ${\cal{F}}^\mathrm{MINERvA}$ considered in Ref.~\cite{MINERvA:2021dhf}, i.e., $0 \le {\cal{F}}^\mathrm{MINERvA} \le 250~\mathrm{keV}$. We provide an analogous comparison for the distributions of the variable ${\cal{F}}$ averaged over the MINERvA medium-energy flux and anticipated DUNE flux~\cite{dune_page} in the following Figs.~\ref{fig:F_spectrum_MINERvA_flux} and~\ref{fig:F_spectrum_DUNE_flux}, respectively. In each case, we observe percent-level  distortions due to $\mathrm{O} \left( \alpha \right)$ radiative corrections. Moreover, there is a significant difference between a naive calculation (applying corrections from Eq.~(\ref{eq:muon_energy_spectrum}) under the assumption of radiation-free kinematics) and the complete calculation which properly accounts for the angular distribution.

\begin{figure}[tp]
\centering
\includegraphics[width=0.79\textwidth]{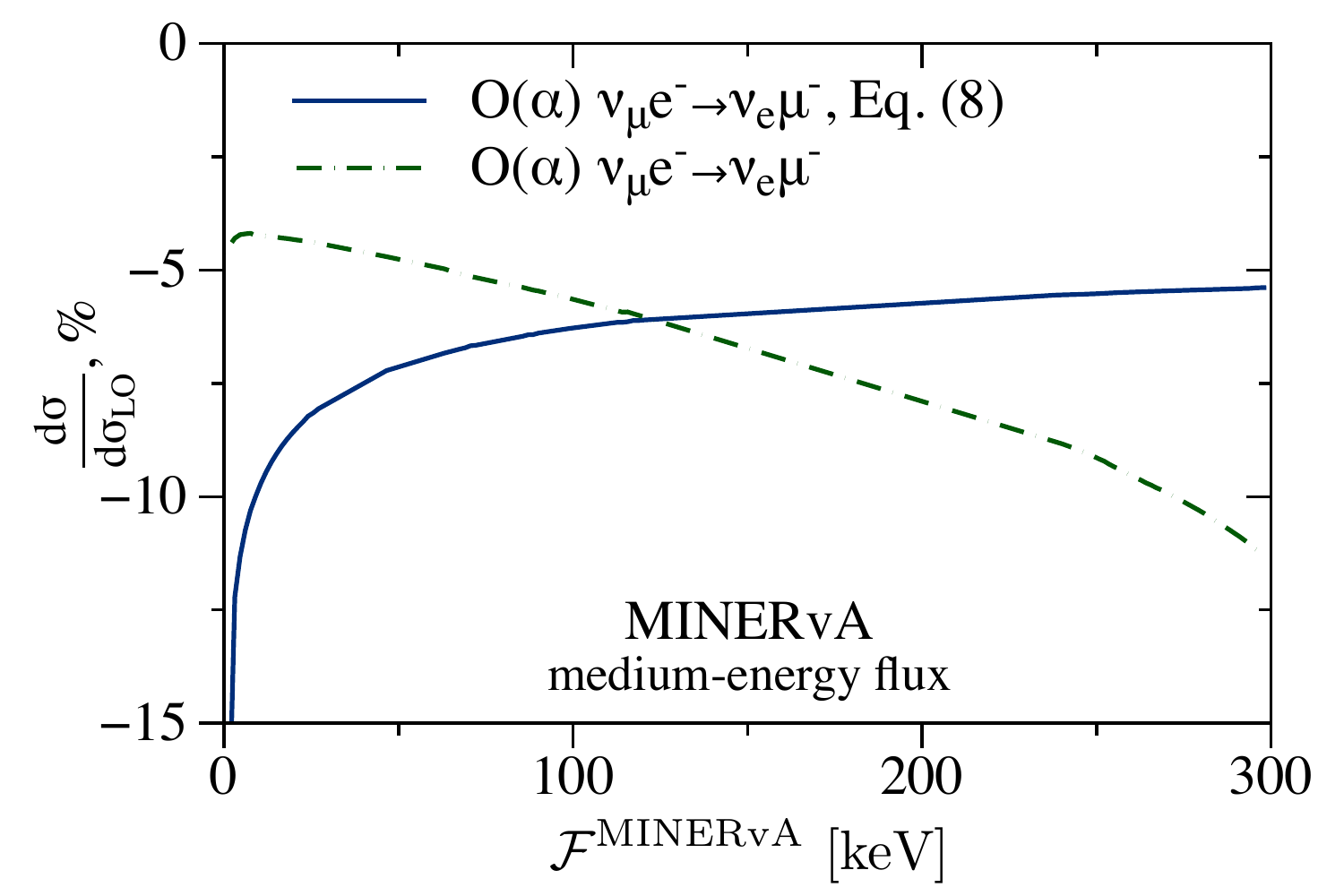}
\caption{Ratios of the $O\left( \alpha \right)$ contribution to the leading-order result for the distribution of the variable ${\cal{F}}^{\mathrm{MINERvA}}$, cf. Eq.~(\ref{eq:variable_F}), averaged over the medium-energy flux of the MINERvA experiment. The $O\left( \alpha \right)$ correction in Eq.(\ref{eq:muon_energy_spectrum}), assuming the kinematics of radiation-free scattering, the blue solid line, is compared to the $O\left( \alpha \right)$ contribution, which is obtained by integrating the appropriate double-differential distribution and adding virtual and soft-photon corrections on top, cf. the green dashed line.
\label{fig:F_spectrum_MINERvA_flux}}
\end{figure}

\begin{figure}[tp]
\centering
\includegraphics[width=0.79\textwidth]{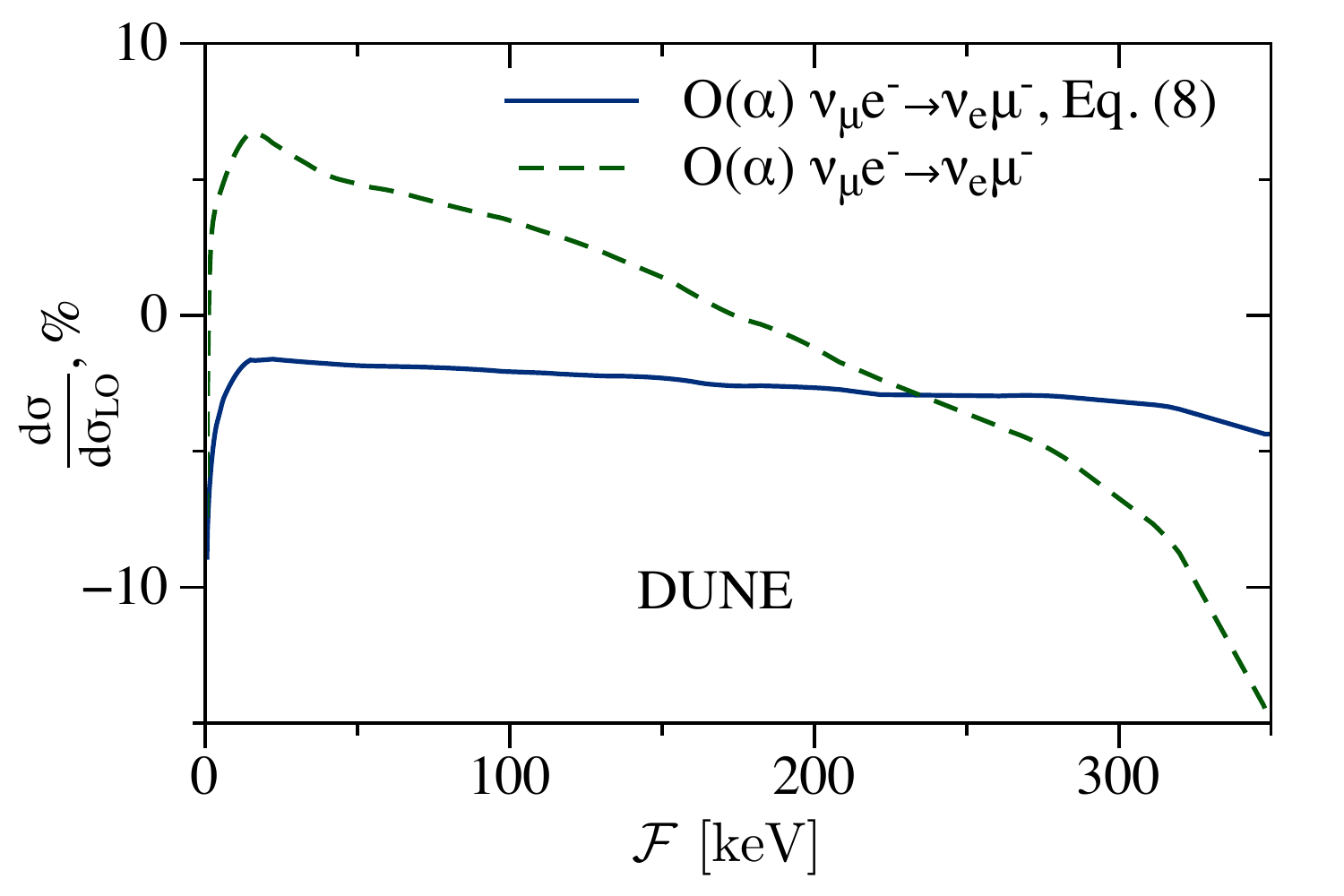}
\caption{Same as Fig.~\ref{fig:F_spectrum_MINERvA_flux} but for the anticipated DUNE flux according to the definition of the variable ${\cal{F}}$ in Eq.~(\ref{eq:variable_F_DUNE}). 
\label{fig:F_spectrum_DUNE_flux}}
\end{figure}

\section{Conclusions and Outlook}
\label{sec7}
 
The goal of this paper is to enable percent-level constraints on the incoming (anti)neutrino fluxes by measuring the inverse muon decay reaction on the atomic electrons. Thus, we performed a study of radiative corrections and various distributions for inverse muon decay. We confirmed an analytical expression for the muon energy spectrum in $\nu_\mu e^- \to {\nu}_e \mu^-$ and presented a new expression for the spectrum in $\bar{\nu}_e e^- \to \bar{{\nu}}_\mu \mu^-$. We provided the following new cross sections: triple-differential distribution in muon energy, muon scattering angle, and photon energy; double-differential distribution in muon energy and muon scattering angle; double-differential distribution in photon energy and photon scattering angle; double-differential distribution in muon energy and photon energy for the dominant muon channel, and total radiative cross section for both channels. 

We investigated the effects of $\mathrm{O} \left(\alpha \right)$ radiative corrections on the muon energy spectrum and on the distribution of the experimentally accessed variable $\cal{F}$. In both cases, the corresponding distortions have the percent-level size. We have clarified the definition for the variable $\cal{F}$ that discriminates between inverse muon decay and other neutrino interactions.  We noted that there is a significant difference between the complete calculation of ${\cal F}$ from the two-dimensional distribution in muon energy and photon energy, and a naive implementation of the muon energy spectrum.

Providing radiative corrections to the inverse muon decay paves the way for percent-level constraints on high-energy tails for neutrino fluxes of modern and future neutrino oscillation and cross-section experiments.

\FloatBarrier

\section*{Acknowledgments}
O.T. thanks Matthias Heller and Marc Vanderhaeghen for technical discussions while working on other projects. This work is supported by the US Department of Energy through the Los Alamos National Laboratory and by LANL’s Laboratory Directed Research and Development (LDRD/PRD) program under project number 20210968PRD4. Los Alamos National Laboratory is operated by Triad National Security, LLC, for the National Nuclear Security Administration of U.S. Department of Energy (Contract No. 89233218CNA000001).  This work was also supported by the U.S. Department of Energy, Office of Science, Office of High Energy Physics, under Awards DE-SC0019095 and DE-SC0008475.  K.S.M. acknowledges support from a Fermilab Intensity Frontier Fellowship during the early stages of this work, and from the University of Rochester's Steven Chu Professorship in Physics. D.R. gratefully acknowledges support from a Cottrell Postdoctoral Fellowship, Research Corporation for Scientific Advancement Award No. 27467 and National Science Foundation Award CHE2039044.   FeynCalc~\cite{Mertig:1990an,Shtabovenko:2016sxi}, LoopTools~\cite{Hahn:1998yk}, Mathematica~\cite{Mathematica}, and DataGraph were extremely useful in this work.

\appendix

\section{Virtual corrections}
\label{sec3}

To evaluate virtual contributions, it is convenient to express vertex corrections as a deviation of the charged-lepton current $\left(\delta \mathrm{J}^\mathrm{L}\right)^\nu$ from the tree-level expression $\left(\mathrm{J}^\mathrm{L}\right)^\nu = \bar{\mu} \left( p_\mu \right) \gamma^\nu \mathrm{P}_\mathrm{L} e \left( p_e \right) $ as
\begin{align}
\left(\delta \mathrm{J}^\mathrm{L}\right)^\nu= e^2 \int \frac{\mathrm{d}^D L }{(2 \pi)^D} \frac{\bar{\mu} \left( p_\mu \right) \gamma^\lambda \left( \slash{p}_\mu - \slash{L} + m_\mu \right)  \gamma^\nu \mathrm{P}_\mathrm{L}  \left( \slash{p}_e - \slash{L} + m_e \right) \gamma^\rho e \left( p_e \right)  }{\left[ (p_\mu - L)^2 - m_\mu^2 \right] \left[(p_e - L)^2 - m_e^2 \right]} \mathrm{\Pi}_{\lambda \rho} \left( L \right), \label{eq:change_of_current}
\end{align}
with the momentum-space photon propagator $\mathrm{\Pi}^{\mu \nu}$:
\begin{align}
\mathrm{\Pi}^{\mu \nu} \left( L \right) = \frac{i}{L^2 - \lambda^2} \left[ - g^{\mu \nu} + \left( 1 - \xi_\gamma \right) \frac{L^{\mu} L^{\nu}}{L^2 - a \xi_\gamma \lambda^2 }\right], \label{eq:photon_propagator}
\end{align}
where the photon mass $\lambda$ regulates the infrared divergence, $\xi_\gamma$ is the photon gauge-fixing parameter, and $a$ is an arbitrary constant. The corresponding field renormalization factors for the external charged leptons are evaluated from the one-loop self-energies as~\cite{Vanderhaeghen:2000ws,Heller:2018ypa}
\begin{align}
Z_\ell &= 1 -  \frac{\alpha}{4\pi} \frac{\xi_\gamma}{\varepsilon} - \frac{\alpha}{4\pi} \left(  \ln \frac{\mu^2}{m_\ell^2} + 2\ln \frac{\lambda^2}{m_\ell^2} + 4 \right) + \frac{\alpha}{4\pi} \left( 1 - \xi_\gamma  \right) \left( \ln \frac{\mu^2}{\lambda^2} + 1 + \frac{a \xi_\gamma \ln  a \xi_\gamma }{1 - a \xi_\gamma }  \right),  \label{eq:charged_lepton_Z_factor}
\end{align}
with the renormalization scale in dimensional regularization $\mu$, where the number of dimensions is $D=4-2 \varepsilon$. Neglecting Lorentz structures whose contractions with the (anti)neutrino current vanish at $m_\nu = 0$ and denoting the ratio of lepton masses as $r = m_e/m_\mu$, the resulting correction to the charged lepton current is expressed as
\begin{align}
\left( \sqrt{Z_e Z_\mu} - 1 \right) \left(\mathrm{J}^\mathrm{L}\right)^\nu +  \left(\delta \mathrm{J}^\mathrm{L}\right)^\nu = \frac{\alpha}{2\pi} \bar{\mu} \left( p_\mu \right) \left( g_M \gamma^\nu - f_2 \frac{p_\mu^\nu+ r p_e^\nu}{2m_\mu r^2} - g^5_M \gamma^\nu \gamma_5 + f^5_2 \frac{p_\mu^\nu - r p_e^\nu}{2m_\mu r^2}   \gamma_5 \right) e \left( p_e \right) , \label{eq:gf}
\end{align}
where the form factors $g_M,~g^5_M,~f_2$, and $f^5_2$ are~\cite{Tomalak:2021lif}
\begin{align}
g^{(5)}_M \left( \eta, r, \beta \right) &=- 1 + \frac{1}{\beta} \left( \frac{1}{2} \left(2 \beta -   \ln \frac{1+\beta}{1-\beta} \right)\ln \frac{2m_e}{\lambda}+ \frac{1}{2} \ln \frac{1+\beta}{1-\beta}  \ln \frac{1+\beta}{ \beta} -  \ln \frac{r \sqrt{1 - \beta} -  \sqrt{1 + \beta} }{  r \sqrt{1 + \beta}  - \sqrt{1 - \beta} } \frac{\ln r}{2} \right. \nonumber \\
&+ \left. \frac{3}{8}  \ln \frac{1+\beta}{1-\beta} + \frac{ \sqrt{1-\beta^2}}{8\eta}   \ln \frac{1+\beta}{1-\beta} + \frac{1}{4}  \ln \frac{1+\beta}{1-\beta}  \ln \frac{2 r  - \left(1 + r^2 \right)\sqrt{1 - \beta^2}}{ 1-\beta} + \frac{\pi^2}{12}  \right. \nonumber \\
&+ \left. \frac{1}{2} \mathrm{Li}_2 \frac{1-\beta}{1+\beta}  - \frac{1}{2} \mathrm{Li}_2 \left( \frac{ \sqrt{1-\beta}}{\sqrt{1+\beta}} r \right) - \frac{1}{2} \mathrm{Li}_2 \left( \frac{\sqrt{1-\beta}}{\sqrt{1+\beta}} \frac{1}{r} \right)     - \frac{5}{16} \ln^2 \frac{1+\beta}{1-\beta} -  \frac{1}{4} \ln^2 r  \right) \nonumber \\
&+ \frac{\sqrt{1-\beta^2} }{8 \beta} \frac{  \left( r + \eta \right)^2 \left( 1 - \eta \sqrt{1-\beta^2} \right) }{2r-\left( 1+ r^2 \right)\sqrt{1-\beta^2}}   \ln \frac{1+  \beta}{1-\beta}  - \frac{12 r - \left( 7 + 5 r^2 \right) \sqrt{1-\beta^2}  }{2r-\left( 1+ r^2 \right)\sqrt{1-\beta^2}} \frac{\ln r}{4} - \ln \frac{2}{r} ,\label{form_factor_g} \\
f^{(5)}_2 \left( \eta, r, \beta \right) &= \frac{r^2}{2} \frac{\sqrt{1-\beta^2}}{\beta} \frac{1-\eta \sqrt{1-\beta^2}}{2r-\left( 1+ r^2 \right)\sqrt{1-\beta^2}} \ln \frac{1+\beta}{1-\beta} -  \frac{r - \eta}{r + \eta}  \frac{r^2 \sqrt{1-\beta^2}}{2r-\left( 1+ r^2 \right)\sqrt{1-\beta^2}} \ln r \,. \label{eq:form_factor_f}
\end{align}
Here $\beta$ is the velocity of the muon in the electron rest frame, $\eta = 1$ for the form factors $g_M,~f_2$ and $\eta = -1$ for the form factors $g^5_M,~f^5_2$. 

The expressions above were presented in the literature using this approach in  Ref.~\cite{Tomalak:2021lif}. Technically equivalent evaluations of physical observables and similar quantities with distinct intermediate steps were performed in Refs.~\cite{Behrends:1955mb,Arbuzov:2001ui,Arbuzov:2004zr}.  Refs.~\cite{Chen:2018dpt,Engel:2018fsb,Anastasiou:2005pn,Engel:2019nfw} have explored higher-order expansions in $\alpha$ for electromagnetic transitions between fermions of different mass.

\section{Real radiation}
\label{sec4}

Inverse muon decay with single photon emission is described by the Bremsstrahlung contribution $\mathrm{T}^{1\gamma}$:
\begin{align}
\mathrm{T}^{1\gamma} = - 2 \sqrt{2} \mathrm{G}_\mathrm{F} i e ~\bar{\nu}_{e}\gamma^\mu \mathrm{P}_\mathrm{L} \nu_{\mu}
  \, \left[  \left( \frac{ {p}^\nu_\mu }{p_\mu \cdot k_\gamma } - \frac{ {p}^\nu_e }{  p_e \cdot k_\gamma }  \right)  \bar{\mu} \gamma_\mu \mathrm{P}_\mathrm{L} e + \frac{1}{2} \bar{\mu}  \left( \frac{ \gamma^\nu \slash{k}_\gamma \gamma_\mu}{  p_\mu \cdot k_\gamma } + \frac{ \gamma_\mu \slash{k}_\gamma \gamma^\nu}{  p_e \cdot k_\gamma } \right)  \mathrm{P}_\mathrm{L} e \right] \varepsilon^\star_\nu, \label{eq:real_radiation}
\end{align}
with the photon polarization four-vector $\varepsilon^\star_\nu$. Let us consider separately the cases of soft and hard photon emission.

\subsection{Soft-photon Bremsstrahlung}
\label{sec4:soft}

The inverse muon decay with radiation of photons of arbitrary small energy cannot be experimentally distinguished from the decay without radiation. All events with photons below some energy cutoff $k_\gamma \le \Delta E$ (in the electron rest frame) must be included in measured observables. The corresponding scattering cross section factorizes in terms of the tree-level result of Eqs.~(\ref{eq:tree_level_inverse_muon_decay_xsection_neutrino}) and (\ref{eq:tree_level_inverse_muon_decay_xsection_antineutrino}) as
\begin{align}
  \mathrm{d} \sigma \left(\nu_\mu e^- \to {\nu}_e \mu^- \gamma,~\bar{\nu}_e e^- \to \bar{{\nu}}_\mu \mu^- \gamma;~k_\gamma \le \Delta E \right) =  \frac{\alpha}{\pi} \delta_s \left( \Delta E \right) \mathrm{d} \sigma_\mathrm{LO} \left(\nu_\mu e^- \to {\nu}_e \mu^-,~\bar{\nu}_e e^- \to \bar{{\nu}}_\mu \mu^-  \right), \label{eq:soft_inverse_muon_decay}
\end{align}
with the universal correction $\delta_s \left( \Delta E \right) $~\cite{Lee:1964jq,Aoki:1980ix,Sarantakos:1982bp,Passera:2000ug,Tomalak:2019ibg}:
\begin{align}
\delta_s \left( \Delta E \right) =  \frac{1}{ \beta}\left(  \mathrm{Li}_2 \frac{1-\beta}{1+\beta} - \frac{\pi^2}{6} \right)-  \frac{2}{\beta} \left( \beta -   \frac{1}{2} \ln \frac{1+\beta}{1-\beta} \right)\ln  \frac{2 \Delta E}{\lambda}  + \frac{1}{2 \beta}  \ln \frac{1+\beta}{1-\beta}\left( 1  +  \ln \frac{ \beta^{-2} \sqrt{1-\beta^2} }{4 \left(1+ \beta \right)^{-1}}   \right)+1\,. \label{eq:soft_result}
\end{align}
This region of the phase space with low-energy photons cancels the infrared-divergent contributions from virtual diagrams. As a result, soft and virtual contributions multiply the tree-level cross sections of Eq.~(\ref{eq:LO_differential}) with infrared-finite factor, i.e., independent of the fictitious photon mass $\lambda$~\cite{Bloch:1937pw,Nakanishi:1958ur,Kinoshita:1962ur,Lee:1964is}, as
\begin{align}
& \frac{\mathrm{d} \sigma \left( \nu_\mu e^- \to {\nu}_e \mu^- \right)  + \mathrm{d} \sigma \left( \nu_\mu e^- \to {\nu}_e \mu^- \gamma \left( k_\gamma \le \Delta E \right) \right)}{\mathrm{d} \sigma_\mathrm{LO} \left( \nu_\mu e^- \to {\nu}_e \mu^-  \right)} =  1 + \frac{\alpha}{\pi} \Bigg\{ g_M + g^5_M + \delta_s \left( \Delta E \right)  \nonumber \\
& + \frac{r m^2_\mu}{4} \frac{\left( p_\mu - p_e \right)^2}{  p_\mu \cdot k_{{\nu}_e} p_e \cdot k_{\nu_\mu} }  \left( g_M - g^5_M \right) - \left( \frac{r^2 m^2_\mu}{4} \frac{\left( p_\mu - p_e \right)^2}{p_\mu \cdot k_{{\nu}_e} p_e \cdot k_{\nu_\mu}} +  \frac{ p_e \cdot k_{{\nu}_e}    }{  p_\mu \cdot k_{{\nu}_e} } \right) \left[ \left( \frac{1+r}{2 r} \right)^2 f_2 + \left( \frac{1-r}{2 r} \right)^2 f^5_2 \right]  \Bigg\}, \label{eq:tree_level_soft_and_virtual_inverse_muon_decay_neutrino}
\end{align}
where we can obtain the contributions in the $\bar{\nu}_e e^- \to \bar{{\nu}}_\mu \mu^-$ reaction by replacing the momenta of neutrinos with the momenta of antineutrinos.

\subsection{Contribution of hard photons}
\label{sec4:hard}

Here we evaluate the contribution of photons above the energy cutoff $k_\gamma \ge \Delta E$ to the muon energy spectrum. Squaring the matrix element of Eq.~(\ref{eq:real_radiation}) for the inverse muon decay $\nu_\mu e^- \to {\nu}_e \mu^-$, we obtain the result in terms of Lorentz invariants as
\begin{align}
\frac{|\mathrm{T}^{1\gamma}|^2 }{e^2 |\mathrm{T}_\mathrm{LO}|^2} &=  - \left( \frac{p_\mu}{p_\mu \cdot k_\gamma} - \frac{p_e}{p_e \cdot k_\gamma} \right)^2 +  \frac{p_\mu \cdot p_e}{p_e \cdot k_\gamma p_\mu \cdot k_\gamma } \left( \frac{k_{{\nu}_e} \cdot k_\gamma }{k_{{\nu}_e} \cdot p_\mu} - \frac{k_{\nu_\mu} \cdot k_\gamma}{k_{\nu_\mu} \cdot p_e} \right) +  \frac{k_{{\nu}_\mu} \cdot k_\gamma}{p_e \cdot k_{{\nu}_\mu} p_e \cdot k_\gamma}  -  \frac{p_e \cdot k_{\nu_e}}{p_\mu \cdot k_{\nu_e} p_e \cdot k_\gamma} \nonumber \\
&+ \frac{1}{p_\mu \cdot k_\gamma} - \frac{1}{p_e \cdot k_\gamma} + \frac{ k_{\nu_e} \cdot k_\gamma}{p_\mu \cdot k_{\nu_e} p_\mu \cdot k_\gamma}  + \frac{p_\mu \cdot k_{{\nu}_\mu}}{p_e \cdot k_{{\nu}_\mu} p_\mu \cdot k_\gamma} + \frac{ k_{\nu_\mu} \cdot k_\gamma }{ \left( p_e \cdot k_\gamma \right)^2 } \frac{m^2_e}{p_e \cdot k_{\nu_\mu}} - \frac{k_{{\nu}_e} \cdot k_\gamma }{ \left( p_\mu \cdot k_\gamma \right)^2 } \frac{m^2_\mu}{p_\mu \cdot k_{{\nu}_e}}, \label{eq:matrix_element_radiative_inverse_muon_decay}
\end{align}
while the result for $\bar{\nu}_e e^- \to \bar{\nu}_\mu \mu^-$ is given by the replacement of neutrino momenta by corresponding antineutrino momenta.

We perform the integration following the technique that was introduced in~\cite{Ram:1967zza} and further developed in~\cite{Tomalak:2019ibg,Tomalak:2021lif,Tomalak:2022xup,Tomalak:2022jhh}. For the inverse muon decay, the implementation is slightly more involved by having two mass scales: the electron mass and the muon mass.

First, we introduce the four-vector $l$: $l = p_e + k_{\nu} - p_{\mu} = \left( l_0,\vec{f} \right)$. Working in the rest frame of atomic electrons, we have for the components of $l$:
\begin{align}
l_0 &= m_e + E_{\nu} - E_{\mu}, \\
f^2 &= |\vec{f}|^2 = E_{\nu}^2 + \beta^2 E^2_\mu - 2 \beta E_{\nu}  E_\mu \cos \theta_\mu. \label{eq:f_components}
\end{align}
Accounting for the conservation of energy and momentum, we obtain
\begin{align}
l^2 &= 2 k_\gamma \left( l_0 - f \cos \gamma \right), \label{eq:matrix_element_radiative_muon}
\end{align}
where $\gamma$ denotes the angle between the photon direction and the vector $\vec{f}$. Using energy and momentum conservation to perform the integration over the final-state neutrino momentum components and the photon energy, we obtain the muon energy spectrum as
\begin{align}
\frac{\mathrm{d} \sigma^{1 \gamma}}{\mathrm{d} E_\mu} &= \int \frac{|\mathrm{T}^{1\gamma}|^2}{256 \pi^4 m_e} \frac{k^2_\gamma f \mathrm{d} f \mathrm{d} \Omega_{k_\gamma}}{E^2_{\nu} \left( l_0^2 - f^2 \right)}. \label{eq:muon_energy_spetrctum_before_integration}
\end{align}
It is convenient to split the phase space into two regions with distinct ranges of integration. There are no restrictions on the photon phase space in the region I: $l^2 \ge 2 \Delta E \left( l_0 - f \cos \gamma \right)$. In this region, the range of kinematic variables is given by
\begin{align}
m_e + \frac{2 \left(\Delta E \right)^2}{m_e - 2 \Delta E} + \frac{\frac{m^2_\mu - m^2_e}{2}}{m_e - 2 \Delta E}&\le E_\mu \le m_e + \frac{2 \left(E_{\nu} - \Delta E \right)^2}{m_e + 2 \left( E_{\nu} - \Delta E \right)} + \frac{\frac{m^2_\mu - m^2_e}{2}}{ m_e + 2 \left( E_{\nu} - \Delta E \right)},  \label{eq:limits_in_region1_Emu} \\
| E_{\nu} - \beta E_\mu | &\le f \le l_0 - 2 \Delta E, \label{eq:limits_in_region1_f} \\
\frac{l_0-f}{2} &\le k_\gamma \le \frac{l_0+f}{2}. \label{eq:limits_in_region1_kgamma}
\end{align}
In the complementary region II: $l^2 \le 2 \Delta E \left( l_0 - f \cos \gamma \right)$ that is close to the kinematics of the radiation-free process, the angle between the photon momentum and the vector $\vec{f}$ is restricted as
\begin{align}
\cos \gamma \ge \frac{1}{f} \left( l_0 - \frac{l^2}{2 \Delta E} \right). \label{eq:region2}
\end{align}
This region contributes a factorizable contribution $\delta_\mathrm{II}$, which adds linearly to $\delta_s \left( \Delta E \right)$ of Eq.~(\ref{eq:tree_level_soft_and_virtual_inverse_muon_decay_neutrino})~\cite{Tomalak:2019ibg}:
\begin{align}
\delta_\mathrm{II} &=   \frac{1}{\beta}\left( \left( \frac{1}{2} +  \ln \frac{\rho \left( 1 + \cos \delta_0 \right) }{4 \beta} \right) \ln \frac{1-\beta}{1+\beta} -\mathrm{Li}_2 \frac{1-\beta}{1+\beta}  - \mathrm{Li}_2 \frac{ \cos \delta_0 -1 }{\cos \delta_0 + 1} +\mathrm{Li}_2 \left(\frac{ \cos \delta_0 -1 }{ \cos \delta_0 + 1} \frac{1+\beta}{1-\beta} \right)  + \frac{\pi^2}{6} \right) \nonumber \\
&+ \ln \frac{1 - \beta \cos \delta_0}{\rho}   - 1, \label{eq:region_2_muon_energy_spectrum}
\end{align}
where the angle $\delta_0$ is given by
\begin{align}
\cos \delta_0 = \frac{E_{\nu}^2 - \beta^2 E_\mu^2 - l_0^2 }{2 \beta E_\mu l_0} \, , \label{eq:angle_delta}
\end{align}
and $\rho = \sqrt{1-\beta^2}$. Only the first term from Eq.~(\ref{eq:matrix_element_radiative_inverse_muon_decay}) contributes in this region. The same term generates the $\Delta E$ dependence after integrating over region I. For the muon energy spectrum including both soft and hard photons, this dependence cancels with the contribution of soft photons from Eq.~(\ref{eq:soft_inverse_muon_decay}). For other terms, we can safely set $\Delta E = 0$ starting from Eq.~(\ref{eq:matrix_element_radiative_inverse_muon_decay}).

\section{Triple-differential distribution}
\label{app:3xsec}

In the following appendixes, we provide analytic expressions for a few unpolarized cross sections of interest for the dominant $\nu_\mu e^- \to \nu_\mu e^-$ channel. The triple-differential cross section with respect to the muon angle, muon energy, and photon energy is given by
\begin{align}
\frac{\mathrm{d} \sigma}{\mathrm{d} E_\mu \mathrm{d} f \mathrm{d} k_\gamma } =  \frac{\mathrm{G}_\mathrm{F}^2}{2 \pi m_e E_\nu^2} \frac{\alpha}{\pi} \mathrm{I}, \label{eq:tripple_differential}
\end{align}
\begin{align}
 \mathrm{I} & = \frac{\rho f^2}{k_\gamma} \frac{\left( k_\gamma + E_\mu \right) \left(l^2 \right)^2 + \left(k_\gamma \left( s + m_\mu^2 \right) - E_\mu l^2 \right) \left( 2 s - m_\mu^2 - m^2_e\right)+ 4 E_\mu E_\nu m_e \left( s - m^2_\mu \right) - 2 k_\gamma s l^2}{\sqrt{d}} \nonumber \\
& - \frac{2 \rho^2 m_\mu^2 m_e \left( s - m^2_\mu \right) E_\nu f^4 \sigma}{d^{3/2}} - \frac{3}{16} \frac{m_e + k_\gamma}{m_e} \frac{\sigma^2}{\rho^2 f^4 k_\gamma^2} + \frac{m_e + k_\gamma}{m_e} \frac{\left( E_\mu^2 + k_\gamma^2 + 4 m_e k_\gamma \right)\left(l^2\right)^2}{4 f^2 k_\gamma^2}  - \frac{2 m_e^2 E_\nu^2 l_0^2}{f^2 k_\gamma^2}  \nonumber \\
& - \frac{m_e + k_\gamma}{m_e} \frac{\left( l^2 - k_\gamma^2 - 2 l_0 k_\gamma \right)\left(m_\mu^2 + m_e^2 \right)^2}{4 f^2 k_\gamma^2} + \frac{2 m_e E_\nu \left( m_e + E_\nu \right) l_0 l^2}{f^2 k_\gamma^2} + \frac{\left( 2 l^2 - 4 E_\nu l_0 + l_0^2 \right) \left( m_\mu^2 - m_e^2 \right)}{f^2} \nonumber \\
& -  \frac{m_e + k_\gamma}{m_e} \frac{\left( m_\mu^2 - m_e^2 - 2 m_e \left( l_0 -3 \left( m_e + E_\nu \right) \right) \right) l^2 -2 \left( E_\nu^2 + 2 E_\nu l_0 -m_\mu^2 \right) m_e^2}{2 f^2 } - \frac{\left( E_\nu^2 - 4 m_e l_0 \right) l^2}{f^2} \nonumber \\
& - \frac{\left( m_e \left( 9 l^2 - 4 m_\mu^2 \right) - 4 l_0 \left( m_\mu^2 + 2 m_e^2 \right)\right) l^2}{4 f^2 k_\gamma^2} + \frac{ \left( l_0 - 2 k_\gamma \right) \left( l_0 - 2 m_e - 2 E_\nu \right) \left( l^2 \right)^2}{2 f^2 k_\gamma^2} - \frac{m_e^2 \left( l^2 - 4m_\mu^2 \right) l^2}{4 f^2 k_\gamma^2}\nonumber \\
& - \frac{\left( \left( l^2 \right)^2 + E_\nu \left( l_0 -m_e - k_\gamma \right) l^2 +2 \left( l_0 +E_\nu \right) k_\gamma^3 + 2 l_0^3 k_\gamma \right) \left( m_\mu^2 - m_e^2 \right)}{2 f^2 k_\gamma^2} - \frac{E_\nu \left( l^2 + k_\gamma l_0 \right) l^2}{f^2 k_\gamma} \nonumber \\
& - \frac{m_e \left( \left( l_0 - k_\gamma \right) \left( l^2 - 4 k_\gamma l_0 \right) + \left( 2 k_\gamma^2 + 6 k_\gamma l_0 - 2 l_0^2 \right) E_\nu  \right) \left( m_\mu^2 - m_e^2 \right)}{2 f^2 k_\gamma^2}  - \frac{l_0 \left( l_0^2 l^2 + 2 m_\mu^2 m_e^2 \right) }{f^2 k_\gamma} \nonumber \\
& + \frac{k_\gamma}{m_e} \frac{\left( \left( l_0^2 -m_\mu^2 - 5 k_\gamma \left( l_0 - E_\nu \right) - 2 \left( E_\nu - k_\gamma \right) l_0 \right) l^2 - \left( E_\nu l_0 - k_\gamma \left( 2 l_0 + 3 E_\nu \right) \right) m_\mu^2 \right) l^2}{2 f^2 k_\gamma^2} \nonumber \\
& + \frac{\left( 2 E_\nu l_0 \left( 3 l^2 - 8 m_e l_0 + 4m_e^2\right) + 4 l_0^2 \left( m_\mu^2 +2 m_e l_0 - 5 m_e^2 \right) +l^2 \left( 2 m_\mu^2 + 7 m_e l_0 \right) \right)E_\nu}{2 f^2 k_\gamma^2}  \nonumber \\
& - \frac{12 E_\nu \left( E_\mu l^2 + m_e l_0 \left( l_0 - E_\nu \right) \right) + 4 m_e l_0 \left( m_e l_0 +m_\mu^2 \right) + m_e E_\nu \left( 3 l^2 - 4 m_e l_0 \right)}{2 f^2}, \label{eq:tripple_differential2}
\end{align}
with the kinematic notations
\begin{align}
\sigma &= \rho \left( E_\nu^2 - f^2 - E_\mu^2 + m_\mu^2 \right) \left(l^2-2 k_{\gamma } l_0 \right) + 4 k_{\gamma } m_\mu f^2  , \\
d &= \beta^2 m_\mu^2 l^2 E_\nu^2 \left(  l^2 + 4 k_{\gamma }^2 - 4 k_{\gamma } l_0\right)  \sin^2 \theta_\mu + \frac{\sigma^2}{4},
\end{align}
and the squared energy in the center-of-mass reference frame $s = m^2_e + 2 m_e E_\nu$.

\section{Double-differential distribution in muon energy and muon angle}
\label{app:2xsec_muon_energy_muon_angle}

Integrating Eq.~(\ref{eq:tripple_differential}) over the photon energy $k_\gamma$, we obtain the double-differential cross section with respect to the recoil muon energy and muon angle. The result is expressed in a similar to elastic neutrino-electron scattering form~\cite{Tomalak:2019ibg}:
\begin{equation}
\frac{\mathrm{d} \sigma}{\mathrm{d} E_\mu \mathrm{d} f} = \frac{\mathrm{G}_\mathrm{F}^2}{\pi E_\nu^2} \frac{m_\mu}{m_e} \frac{\alpha}{\pi}\left( a  + b \frac{f}{l^2_0 - f^2} \ln  \frac{1+\beta}{1-\beta} + c \ln \frac{l_0+f}{l_0-f} +  \tilde{d} \ln \frac{l_0 - \beta f \cos \delta - \sqrt{g}}{l_0 - \beta f \cos \delta + \sqrt{g}} \right), \label{eq:muon_2Dspectrum}
\end{equation}
with $ g = \left( f \cos \delta - \beta l_0\right)^2 + \rho^2  f^2 \sin^2 \delta $ and the angle $\delta$ between vectors $\vec{f}$ and $\vec{p}_\mu$:
\begin{equation}
\cos \delta = \frac{E_\nu^2 - \beta^2 E_\mu^2 - f^2 }{2 \beta E_\mu f}. \label{eq:angle_delta2}
\end{equation}
The coefficients $a,~b,~c,$ and $\tilde{d}$ are given by
\begin{align}
a &= \frac{\beta \cos \delta }{\rho} \left( m^2_\mu - m^2_e \right) + \frac{f}{m_\mu} \left( s - m^2_\mu - m^2_e - \frac{2\left( s - m^2_\mu \right)\left( s - m^2_e \right)}{l^2}\right)+ \left( f + \frac{10  m_\mu \beta \cos \delta }{\rho}  \right) \frac{l_0 - \beta f \cos \delta }{2 \rho}\nonumber \\
&+ \left( 1- \frac{l_0}{4 m_e} \right) \frac{m_\mu}{f}  \frac{\beta^2}{\rho^2} \left( 1 - 3 \cos^2 \delta \right) l^2 + \frac{m^2_\mu + m^2_e}{m_e} \frac{f - \beta l_0 \cos \delta}{2 \rho}  -  \frac{m_\mu}{m_e} \frac{\beta \cos \delta}{\rho} \frac{2l^2_0 + f^2}{\rho}  \nonumber \\
&+  \frac{3 l_0 f}{2} \frac{m_\mu}{m_e} \frac{1+\beta^2 \cos^2 \delta}{\rho^2}, \, \\
b &= \frac{2 m_\mu l^2_0}{\beta \rho^2}  + \frac{m_e E_\nu}{\beta  m_\mu} \left( s - m^2_\mu \right) + \frac{m^2_\mu-m^2_e}{\beta} \frac{l_0 - \beta f \cos \delta }{\rho} - 4 m_\mu f \frac{l_0 -\frac{1}{2} \beta f \cos \delta }{\rho} \frac{\cos \delta }{\rho}, \\
c &=\frac{m_\mu}{2} \left( l_0 + \frac{m_\mu}{\rho} \right) - \frac{s^2 + m_e^3 E_\nu}{2 m_\mu m_e} - \left( 1 + \frac{2 \beta \cos \delta}{\rho} \frac{m_\mu}{f} - \frac{m^2_\mu}{f^2}  \frac{\beta^2}{\rho^2} \frac{ 1 - 3 \cos^2 \delta}{2} \right) \frac{l^2\left(l^2  -4 m_e l_0 \right)}{4 m_\mu m_e}  \nonumber \\
&+ \left( 1 + \frac{\beta \cos \delta}{\rho} \frac{m_\mu}{f} \right) \left(\frac{1}{\rho} - \frac{m^2_\mu + m^2_e}{4 m_\mu m_e}\right) l^2 - \frac{\beta \cos \delta}{2\rho} \left(\frac{m_\mu}{f} \frac{l^2}{\rho} + \frac{l_0 \left( 2 s - m^2_\mu - m^2_e \right)}{f} \right), \\
\tilde{d} &= \frac{\sqrt{g}}{\rho} f + \frac{\rho}{\sqrt{g}} f \left( s - m_\mu^2 + \frac{m_e E_\nu}{2 m_\mu^2} \left(  s + m_\mu^2 \right) - m_\mu \frac{l_0 - \beta f \cos \delta }{\rho} \right).
\end{align}

\section{Double-differential distribution in photon energy and photon angle}
\label{app:2xsec_photon_energy_photon_angle}

To study cross sections with respect to the photon kinematics, we introduce an ancillary four-vector $\overline{l} $: $\overline{l} = k_\nu + p_e - k_\gamma = \left( \overline{l}_0, \vec{\overline{f}} \right)$. In the laboratory frame, this vector can be expressed as
\begin{align}
\overline{l}_0 &= m_e + E_\nu - k_\gamma, \\
\overline{f}^2 &= |\vec{\overline{f}}|^2 = E_\nu^2 + k^2_\gamma -2 E_\nu k_\gamma \cos \theta_\gamma, \label{eq:photon_kinematics}
\end{align}
with the photon scattering angle $\theta_\gamma$.

The double-differential cross section with respect to the photon angle and photon energy is given by
\begin{align}
\frac{\mathrm{d} \sigma}{ \mathrm{d} k_\gamma \mathrm{d} \overline{f}} = \frac{\mathrm{G}_\mathrm{F}^2}{\pi E_\nu} \frac{\alpha}{\pi} \left( \tilde{a} \left( \overline{l}^2 - m_\mu^2 \right) + \tilde{b} \ln \frac{m_\mu^2}{\overline{l}^2}\right) \frac{\overline{f} }{\left( \overline{l}^2 - s \right)^2},
\label{eq:photon_2D_spectrum}
\end{align}
with coefficients $\tilde{a}$ and $\tilde{b}$:
\begin{align}
\tilde{a} &= -\frac{m_e \left( 2  \overline{l}_0 - m_e \right)}{\overline{l}^2}  \left( s + 2 m^2_\mu + \left( s - m^2_\mu - m^2_e \right) \frac{k_\gamma}{E_\nu} + m^2_\mu \left( \frac{\left(   \overline{l}_0 - m_e \right)^2}{E_\nu k_\gamma} + \frac{m_e \left( s - m_e k_\gamma \right)}{4 E_\nu k^2_\gamma} \right)  \right)  \nonumber \\
& -\frac{\overline{l}^2 \left( \overline{l}^2 - m_\mu^2 \right)}{4 E_\nu m_e k_\gamma} \left( \frac{\overline{l}^2 - 2 \overline{l}_0 m_e }{m_e} + \frac{\overline{l}^2 - 4 \overline{l}_0 m_e}{k_\gamma} \right) + \frac{m_e s^2}{4 E_\nu k_\gamma^2} - 3 \left( s - m^2_\mu \right) + s \left( \frac{E_\nu}{k_\gamma} + \frac{m_e}{2E_\nu} \left( 1 - \frac{3}{2} \frac{m_e}{k_\gamma} \right) \right)  \nonumber \\
&+ \frac{ \overline{l}^2 - m_\mu^2}{E_\nu} \left( k_\gamma - 3 E_\nu- \frac{3}{4} \frac{ \left( \overline{l}_0 - m_e \right)^2}{k_\gamma} + m_e \frac{m^2_e - \overline{l}_0 \left(2 k_\gamma + 5 \overline{l}_0  \right)}{4 k^2_\gamma} + \frac{\overline{f}^2 \left( m_e - k_\gamma \right)}{4k^2_\gamma}\right)  -\frac{4 m_e^2 k^2_\gamma }{\overline{l}^2}, \\
\tilde{b} &= - \left( \overline{l}^2 - s \right)^2 - 2 \left( s - m^2_\mu \right) \left( \overline{l}^2 + m_\mu^2 \right).
\label{eq:photon_2D_spectrum_coefficients}
\end{align}

\section{Photon energy spectrum}
\label{app:2xsec_photon_energy}

Integrating Eq.~(\ref{eq:photon_2D_spectrum_coefficients}) over the photon scattering angle, we obtain the photon energy spectrum for the photon energy $k_\gamma \ge \frac{m_e}{2} - \frac{m^2_\mu}{2 \left( m_e + 2 E_\nu \right)}$, when the photon scatters in the cone around the forward direction:
\begin{align}
\frac{\mathrm{d} \sigma}{ \mathrm{d} k_\gamma} &= \frac{\mathrm{G}_\mathrm{F}^2}{\pi E_\nu} \frac{\alpha}{\pi} \left( \overline{a} + \overline{b} \ln \frac{2m_e k_\gamma}{s - m^2_\mu} + \overline{c}  \ln \frac{s - 2 m_e k_\gamma}{m^2_\mu} - \overline{d} \ln \frac{2 k_\gamma}{2 E_\nu + m_e} \ln \frac{s - 2 m_e k_\gamma}{m^2_\mu} \right. \nonumber \\
& \left. + \overline{d} \sum \limits_{\sigma_1, \sigma_2 = \pm} {\cal{\Re}} \left( \mathrm{Li}_2 \frac{\overline{l}_0 + \sigma_1 \sqrt{ \overline{l}_0^2- m^2_\mu}}{\overline{l}_0 + \sigma_2 \sqrt{ \left( \overline{l}_0- m_\mu \right)^2 - 2 m_e k_\gamma}} - \mathrm{Li}_2 \frac{\overline{l}_0 + \sigma_1 \left( \overline{l}_0- m_e \right)}{\overline{l}_0 + \sigma_2 \sqrt{ \left( \overline{l}_0- m_\mu \right)^2 - 2 m_e k_\gamma}} \right)\right),
\label{eq:photon_1D_spectrum}
\end{align}
with coefficients $\overline{a},~\overline{b},~\overline{c},$ and $\overline{d}$ in Eq.~(\ref{eq:photon_1D_spectrum}):
\begin{align}
\overline{a} &= \frac{s-m^2_\mu - 2 m_e k_\gamma}{12 m_e E_\nu} \left( m_e k_\gamma - \frac{20 s - 5 m^2_\mu - 14 m^2_e}{2} - \frac{53 s^2 - 38 m^2_e s - m^2_\mu \left( 49 s - 26 m^2_e \right) + 2 m^4_\mu - 6 m^4_e}{4 m_e k_\gamma} \right) \nonumber \\
& -\frac{s-m^2_\mu - 2 m_e k_\gamma}{12 m_e E_\nu}  \frac{s \left( 2 s - 3 m^2_e \right) - m^2_\mu \left( 4 s - 9 m^2_e \right) + 2 m^4_\mu}{4 k^2_\gamma}, \\
\overline{b} &=  - \left( s - m_\mu^2 \right)^2 \left( \frac{m_e + E_\nu}{k_\gamma s}  + \frac{2}{s - m^2_\mu} + \frac{1}{s - m^2_e} \right) ,  \\
\overline{c} &=\frac{m^2_\mu}{2} + \frac{s - 2 m_e k_\gamma}{2} \left( 1 + \frac{s}{m_e k_\gamma} - \frac{m^4_\mu}{m^2_e k^2_\gamma} \frac{k_\gamma s - m^3_e}{4 E_\nu s} \right),  \\
\overline{d} &= - \left( s - m_\mu^2 \right).
\label{eq:photon_1D_spectrum_coefficients}
\end{align}
For smaller energies $k_\gamma \le \frac{m_e}{2} - \frac{m^2_\mu}{2 \left( m_e + 2 E_\nu \right)} < \frac{m_e}{2}$, when there are no restrictions on the photon scattering angle, the photon energy spectrum is given by
\begin{align}
\frac{\mathrm{d} \sigma}{ \mathrm{d} k_\gamma} &= \frac{\mathrm{G}_\mathrm{F}^2}{\pi E_\nu} \frac{\alpha}{\pi} \left( \overline{e} - \overline{b} \ln \frac{s}{m^2_e} + \overline{f}  \ln \frac{s - 2 m_e k_\gamma}{m^2_\mu}   - \overline{d} \left( \ln \frac{2 k_\gamma}{2 E_\nu + m_e} \ln \frac{s - 2 m_e k_\gamma}{m^2_\mu}  - \ln \frac{2 k_\gamma}{m_e} \ln \frac{\left( 1 - \frac{2 k_\gamma}{m_e} \right)s}{m^2_\mu} \right)\right. \nonumber \\
& \left. + \overline{g}  \ln \frac{1 - \frac{2 m_e k_\gamma}{s}}{1 - \frac{2 k_\gamma}{m_e} } + \overline{d} \sum \limits_{\sigma_1, \sigma_2 = \pm} {\cal{\Re}} \left( \mathrm{Li}_2 \frac{\overline{l}_0 + \sigma_1\left( \overline{l}_0- m_e + 2 k_\gamma \right)}{\overline{l}_0 + \sigma_2 \sqrt{ \left( \overline{l}_0- m_\mu \right)^2 - 2 m_e k_\gamma}} - \mathrm{Li}_2 \frac{\overline{l}_0 + \sigma_1 \left( \overline{l}_0- m_e \right)}{\overline{l}_0 + \sigma_2 \sqrt{ \left( \overline{l}_0- m_\mu \right)^2 - 2 m_e k_\gamma}} \right)\right),
\label{eq:photon_1D_spectrum2}
\end{align}
with coefficients $\overline{e},~\overline{f}$, and $\overline{g}$ in Eq.~(\ref{eq:photon_1D_spectrum2}):
\begin{align}
\overline{e} &= - \frac{m^4_\mu}{2m_e^2}  - 2 \left( 1 + \frac{E_\nu}{m_e} \right) \left( s - m^2_\mu - \frac{m^2_e}{2} \right) - 2 k^2_\gamma \frac{E_\nu}{m_e} \left( 1 + \frac{4}{3} \frac{E_\nu}{m_e} \right)- E_\nu k_\gamma \left( 3 + 2 \frac{m^2_\mu}{m^2_e} - \frac{10}{3}  \frac{E_\nu}{m_e} \right) \nonumber \\
&- \frac{m^4_\mu m_e + 8 E_\nu \left( s - m^2_\mu \right)^2}{2 k_\gamma s},  \\
\overline{f} &=  \frac{E_\nu}{k_\gamma} \left( s + 2 k^2_\gamma - \frac{m^4_\mu}{s} \right), \\
\overline{g} &= \frac{\left( 2 k_\gamma - m_e \right) \left( \left( m_e + k_\gamma \right) \left( m^4_\mu - 4 E_\nu k_\gamma s \right) - 2 m_e m^4_\mu \right) + 2 m^2_\mu E_\nu k_\gamma \left( m^2_\mu + 2 m_e k_\gamma \right)}{8 m_e E_\nu k^2_\gamma}.
\label{eq:photon_1D_spectrum_coefficients2}
\end{align}

Please note that the expressions in Appendix G of Ref.~\cite{Tomalak:2019ibg} are valid only for the photon energies $k_\gamma \ge m_e E_\nu / \left( m_e + 2 E_\nu \right)$. Below this energy, they should be modified as (in notations of Ref.~\cite{Tomalak:2019ibg})
\begin{align}
 \tilde{\mathrm{I}}_i &\to \frac{\pi^2}{\omega^3} \mathrm{d} k_\gamma \Bigg[ e_i -   b_i \ln \frac{s}{m^2} + f_i \ln \frac{s - 2 m k_\gamma}{m^2} - d_i \left( \ln \frac{2 k_\gamma}{2\omega + m} \ln \frac{2 \bar{l}_0 - m}{m} - \ln \frac{2 k_\gamma}{m} \ln \frac{\left( 1 - \frac{2 k_\gamma}{m} \right)s}{m^2} \right) \Bigg. \nonumber \\
 &+ \Bigg. g_i  \ln \frac{s - 2 m k_\gamma}{\left( 1 - \frac{2 k_\gamma}{m} \right)s}  + d_i   \sum \limits_{\sigma_1,~\sigma_2 =\pm} \Re \Bigg( \mathrm{Li}_2 \frac{\bar{l}_0+ \sigma_1\left( \overline{l}_0- m + 2 k_\gamma \right)}{\bar{l}_0+ \sigma_2 \sqrt{\left( \bar{l}_0-m\right)^2 - 2 m k_\gamma}} -  \mathrm{Li}_2 \frac{\bar{l}_0 + \sigma_1\left(\bar{l}_0-m\right)}{\bar{l}_0+ \sigma_2 \sqrt{\left(\bar{l}_0-m\right)^2 - 2 m k_\gamma}}  \Bigg) \Bigg],
\label{eq:photon_spectrum_non_universal_part}
\end{align}
with coefficients $e_i,~f_i$, and $g_i$ in Eq.~(\ref{eq:photon_spectrum_non_universal_part}):
\begin{align}
e_\mathrm{L} &= \omega \left( \frac{5}{6}\frac{k_\gamma \omega \left( 2 \omega - 3 m \right)}{m^2} - \frac{k^2_\gamma \omega \left( 4 \omega + 3 m \right)}{3 m^3} - \frac{8 \omega^2 + 6 m \omega - m^2}{4m} - \frac{m \left( 32 \omega^3 + m^3 \right)}{4 k_\gamma s} \right),  \\
e_\mathrm{R} &= - \frac{19}{144} \frac{m^9}{k_\gamma s^3} - \frac{m^7 \left( m - 40 k_\gamma \right)}{96 k^2_\gamma s^2} + \frac{m^3 \left( m - 2 k_\gamma \right)}{48 \left( 2 \omega - 2 k_\gamma - m \right)^2} + \frac{m^2 \left( 24 k^4_\gamma - 16 k^3_\gamma m+ 2 k^2_\gamma m^2 - 4 k_\gamma m^3 + m^4\right)}{192 k^3_\gamma  \left( 2 \omega - 2 k_\gamma - m \right)} \nonumber \\
&+ \frac{m^3 \left( 48 k^4_\gamma - 168 k^3_\gamma m - 210 k^2_\gamma m^2 + 4 k_\gamma m^3 - m^4\right)}{192 k^3_\gamma  s} + \frac{m \left( 36 k^3_\gamma - 153 k^2_\gamma m - 49 k_\gamma m^2 + 59 m^3 \right)}{72 k_\gamma \left( m - k_\gamma \right)} \nonumber \\
&- \frac{k^2_\gamma \omega \left( 4 \omega^2 - 3 m \omega + 6 m^2 \right)}{9 m^3} +\frac{k_\gamma \omega \left( 10 \omega^2 - 81 m \omega -132 m^2 \right)}{18 m^2} - \frac{\omega \left( 4 \omega^2 + 52 m \omega + 67 m^2 \right)}{6 m} \nonumber \\
&-\frac{\omega \left( 68 \omega^2 + 48 m \omega +51 m^2\right)}{46 k_\gamma} ,  \\
e^\mathrm{L}_\mathrm{R} &= \omega \left( \frac{3}{2} \omega - 2 m + k_\gamma \left( 3 \frac{\omega}{m} - 2 \right) + \frac{m^3 \left( 24 \omega^3 - 16 m \omega^2 - 14 m^2 \omega - 3 m^3 \right)}{4 k_\gamma s^2}\right) , \\
b^\mathrm{L}_\mathrm{R} &\to 2 b^\mathrm{L}_\mathrm{R} +\frac{2 k_\gamma \omega^2}{m} , \\
f_\mathrm{L}    &=\frac{k_\gamma \omega^2}{m} + \frac{2 \omega^3 m \left( m + \omega \right)}{k_\gamma s} , \\
f_\mathrm{R} &=\omega \left( \frac{k_\gamma}{s} \left( 2 \omega^2+ 6 m \omega +3 m^2 \right) + \frac{2m^2\left( 3 \omega^2+ 16 m \omega \left( m + \omega \right) + 4 m^3 \right)}{s^2} \right)  \nonumber \\
&+ \frac{2}{3}\frac{m^3 \omega \left( m + \omega \right) \left( 4 \omega^4 + 14 m \omega^3 + 25 m^2 \omega^2 + 15 m^3 \omega + 3 m^4\right)}{k_\gamma s^3}, \\
f^\mathrm{L}_\mathrm{R} &\to  - b^\mathrm{L}_\mathrm{R}- \frac{2 k_\gamma \omega^2}{m},  \\
g_\mathrm{L}    &=\frac{\left( m - 2 k_\gamma \right) \left( 8 \omega^2 k_\gamma \left( m + k_\gamma \right) + m^3 \left( m - k_\gamma \right) \right)+2m k_\gamma  \omega \left( 3 m^2 - 4 k^2_\gamma \right)}{16 m k^2_\gamma} ,  \\
g_\mathrm{R} &= \frac{m^4}{48 k_\gamma^2} - \frac{k_\gamma^2}{3}-\frac{2 \omega^2 + 14 m \omega + 11 m^2}{4} - k_\gamma \left(\frac{\omega^2}{m}+\frac{7}{2} \omega +\frac{15}{4}  m \right) +\frac{m \left( 24 \omega^2 + 132 m \omega + 109 m^2\right)}{48 k_\gamma}, \\
g^\mathrm{L}_\mathrm{R} &=\frac{1}{2} m \left( m - 2 \omega \right) - \frac{3}{16} \frac{m^4}{k^2_\gamma} + k_\gamma \left( m - \frac{2 \omega^2}{m}\right) + \frac{m \left( 4 \omega^2 + 8 m \omega -m^2\right)}{8 k_\gamma},
\end{align}
with the soft-photon limit for the small electron mass $m \ll \omega$:
\begin{align}
    \tilde{\mathrm{I}}_\mathrm{L} & = -\frac{\pi^2}{k_\gamma} \left( 4 + 2 \ln \frac{m}{2 \omega} \right) , \\
    \tilde{\mathrm{I}}_\mathrm{R} & = -\frac{\pi^2}{k_\gamma} \left( \frac{17}{9} + \frac{2}{3} \ln \frac{m}{2 \omega} \right) , \\
    \tilde{\mathrm{I}}^\mathrm{L}_\mathrm{R} & = \frac{m}{\omega}\frac{\pi^2}{k_\gamma} \left( \frac{3}{2} + \ln \frac{m}{2 \omega} \right) .
\end{align}

\bibliography{IMD}{}

\end{document}